\documentclass[11pt,a4paper]{article}
\pdfoutput=1
\usepackage{jheppub}

\usepackage{multirow, graphicx,amssymb,url,mathrsfs,amsmath}
\usepackage{wrapfig,boxedminipage,setspace,subfigure,epsfig}
\usepackage{amsxtra,amstext,latexsym,dsfont,amsfonts}
\usepackage{color,eucal}
\usepackage[dvipsnames]{xcolor}
\usepackage{float}
\usepackage{slashed,comment}
\usepackage{kotex}

\newcommand{\dd}{\mathrm{d}}

\title{The breakdown of magneto-hydrodynamics near AdS$_2$ fixed point and energy diffusion bound}

\author[a,b]{Hyun-Sik Jeong}
\author[c]{Keun-Young Kim,}
\author[a,b]{and Ya-Wen Sun}

\emailAdd{hyunsik@ucas.ac.cn}
\emailAdd{fortoe@gist.ac.kr}
\emailAdd{yawen.sun@ucas.ac.cn}

\affiliation[a]{School of physics, University of Chinese Academy of Sciences, Beijing 100049, China}
\affiliation[b]{Kavli Institute for Theoretical Sciences, University of Chinese Academy of Sciences, Beijing 100049, China}
\affiliation[c]{School of Physics and Chemistry, Gwangju Institute of Science and Technology, \\
123 Cheomdan-gwagiro, Gwangju 61005, Korea}

\abstract{
We investigate the breakdown of magneto-hydrodynamics at low temperature ($T$) with black holes whose extremal geometry is AdS$_2\times$R$^2$. The breakdown is identified by the equilibration scales ($\omega_{\text{eq}}, k_{\text{eq}}$) defined as the collision point between the diffusive hydrodynamic mode and the longest-lived non-hydrodynamic mode. 
We show ($\omega_{\text{eq}}, k_{\text{eq}}$) at low $T$ is determined by the diffusion constant $D$ and the scaling dimension $\Delta(0)$ of an infra-red operator: $\omega_{\text{eq}} = 2\pi T \Delta(0), \, k_{\text{eq}}^2 = \omega_{\text{eq}}/D$, where $\Delta(0)=1$ in the presence of magnetic fields. For the purpose of comparison, we have analytically shown $\Delta(0)=2$ for the axion model independent of the translational symmetry breaking pattern (explicit or spontaneous), which is complementary to previous numerical results. {Our results support the conjectured universal upper bound of the energy diffusion $D  \,\le\,  \omega_{\text{eq}}/k_{\text{eq}}^2 \,:=\, v_{\text{eq}}^2 \, \tau_{\text{eq}}$ where $v_{\text{eq}}:= \omega_{\text{eq}}/k_{\text{eq}}$ and $\tau_{\text{eq}}:=\omega_{\text{eq}}^{-1}$ are the velocity and the timescale associated to equilibration, implying that the breakdown of hydrodynamics sets the upper bound of the diffusion constant $D$ at low $T$.}

}
\begin{document}
\maketitle

%%%%%%%%%%%%%%%%%%%%%%%%%%%%%%%%
%    Section: Introduction
%%%%%%%%%%%%%%%%%%%%%%%%%%%%%%%%
\section{Introduction}

Hydrodynamics is an effective theory for describing near-equilibrium phenomena at late time and large distance~\cite{LLDM87}.
It is constructed by conservation equations together with constitutive relations in a gradient expansion around equilibrium: in the Fourier space, the expansion is in the frequency $\omega$ and momentum $k$.
The gradient expansion reflects that hydrodynamics is an effective description of the system at length scales larger than the characteristic length of the system and, at late times, the dynamics of the system back to equilibrium is dominated by the longest-lived mode referred to as hydrodynamic modes.

One may expect that, as higher orders are considered in the expansion, hydrodynamics will be a more improved description for the near equilibrium physics. For instance, one can schematically write the constitutive relation for the stress tensor as
\begin{equation}\label{}
\begin{split}
T^{\mu\nu} = O(\phi) + O(\nabla\phi)  + \dots\,,
\end{split}
\end{equation}
where $\phi$ denotes the hydrodynamic variables (such as temperature, fluid velocity).
The leading order $O(\phi)$ gives rise to the perfect fluid and it is improved by the next order expansion $O(\nabla\phi)$ to contain the viscous effect: the viscous fluid.

As in other perturbative expansions, one may ask for the convergence of the hydrodynamic expansion such as ``does the hydrodynamic expansion indeed converge?"
In recent years, regarding this question, the breakdown of hydrodynamics has been investigated in~\cite{Withers:2018srf,Grozdanov:2019kge,Grozdanov:2019uhi,Heller:2020hnq,Heller:2020uuy,Heller:2013fn,Abbasi:2020ykq,Jansen:2020hfd,Grozdanov:2020koi,Choi:2020tdj,Arean:2020eus,Wu:2021mkk,Grozdanov:2021gzh}.

In particular, using holography (or the gauge/gravity duality)~\cite{Maldacena:1997re,Witten:1998qj,Gubser:1998bc}, one can characterize the breakdown of hydrodynamics by the equilibration scales ($\omega_{\text{eq}}, k_{\text{eq}}$) defined as the collision point in a complex ($\omega, k$) space between the hydrodynamic mode and the first-non hydrodynamic mode~\cite{Arean:2020eus,Wu:2021mkk}.
At such scales, the lifetime of non-hydrodynamic mode is comparable to the one of the hydrodynamic mode so that the dynamics of the system can no longer be dominated just by a hydrodynamic mode.

In this work, following previous studies~\cite{Arean:2020eus,Wu:2021mkk}, we further study the breakdown of hydrodynamics in the low temperature ($T$) where the black hole has an extremal AdS$_2\times$ R$^2$ geometry\footnote{Low temperature states dual to the AdS$_2\times$R$^2$ geometry are related to the Sachdev-Ye-Kitaev (SYK)-like models in strange metals. One can see the breakdown of hydrodynamics of such black holes gives the consistent results with SYK model in \cite{Arean:2020eus}.}. 
In particular, we will see if the following results \eqref{RES1}-\eqref{RES2} found in previous studies still hold at finite magnetic fields, i.e., we study the breakdown of magneto-hydrodynamics in holography~\cite{Buchbinder:2008dc,Buchbinder:2009aa,Hansen:2009xe,Buchbinder:2009mk,Hartnoll:2007ip,Hansen:2008tq,Hartnoll:2007ih,Hernandez:2017mch,Baggioli:2021ujk}\footnote{{For the recent development of magneto-hydrodynamics, see also \cite{Amoretti:2021fch,Amoretti:2020mkp,Amoretti:2019buu} and references therein.}} .
\paragraph{Result 1:}
\begin{equation}\label{RES1}
\begin{split}
\omega_{\text{eq}} \,\rightarrow\, 2\pi T \Delta(0)  \quad  \text{as}\quad  T\rightarrow 0  \,,
\end{split}
\end{equation}
where $\Delta(0)$ is the infra-red scaling dimension. This result shows the non-hydrodynamic mode is related to the infra-red Green's function so that the breakdown at low $T$ is associated to the emergence of the  AdS$_2\times$R$^2$ geometry.

\paragraph{Result 2:}
\begin{equation}\label{RES2}
\begin{split}
k_{\text{eq}}^2 \,\rightarrow\, \frac{\omega_{\text{eq}}}{D}  \quad  \text{as}\quad  T\rightarrow 0  \,,
\end{split}
\end{equation}
which reflects the fact that, at low $T$, the hydrodynamic dispersion with diffusion constant $D$ at the quadratic order \eqref{SD2} becomes a good approximation even at ($\omega_{\text{eq}}$, $k_{\text{eq}}$)\footnote{This may imply the radius of convergence could be determined by the leading order in the hydrodynamic expansion for the AdS$_2$ fixed point~\cite{Arean:2020eus}.}. 
From these results, one may simply determine the scales at which hydrodynamics break down from the diffusion constant $D$ and the scaling dimension $\Delta(0)$.

{
Rearranging \eqref{RES2}, one can notice that the diffusion constant $D$ at low $T$ can be determined by the breakdown of hydrodynamics ($\omega_{\text{eq}}$, $k_{\text{eq}}$) and can be expressed further as follows
\begin{equation}\label{RES3new}
\begin{split}
D \,\,\rightarrow\,\, \frac{\omega_{\text{eq}}}{k_{\text{eq}}^2} \,\,:=\,\, v_{\text{eq}}^2 \, \tau_{\text{eq}}  \quad  \text{as}\quad  T\rightarrow 0  \,,
\end{split}
\end{equation}
where $v_{\text{eq}}:= \omega_{\text{eq}}/k_{\text{eq}}$ and $\tau_{\text{eq}}:=\omega_{\text{eq}}^{-1}$ are the velocity and the timescale associated with  equilibration~\cite{Arean:2020eus,Wu:2021mkk}.
In this velocity and time scales, it turned out \eqref{RES3new} corresponds to the upper bound of $D$~\cite{Arean:2020eus,Wu:2021mkk}, i.e.,
\begin{equation}\label{RES4new}
\begin{split}
D  \,\le\,  v_{\text{eq}}^2 \, \tau_{\text{eq}}  \,,
\end{split}
\end{equation}
where the upper bound (an equality) is approached at low $T$ \eqref{RES3new}.
Thus, the breakdown of hydrodynamics can be used to set the upper bound of the diffusion constant $D$.
In this work, in addition to \eqref{RES1}-\eqref{RES2}, we will also study if the conjectured upper bound \eqref{RES4new} still hold at finite magnetic fields.
}

Last but not least, our work is related to one of future works proposed in \cite{Arean:2020eus}: the investigation of \eqref{RES1}-\eqref{RES4new} with magnetic fields. They studied the breakdown of the energy diffusive hydrodynamics with axion charge (or chemical potential), and found $\Delta(0)=2$ for those cases. We will study if \eqref{RES1}-\eqref{RES4new} continues to hold for AdS$_2$ fixed points supported by a different hierarchy of scales for the energy diffusion (i.e., magnetic fields) and identify a corresponding $\Delta(0)$.

{
Furthermore, we will present how to obtain $\Delta(0)$ analytically using the infra-red Green's function at finite magnetic fields. Using the same procedure, for the purpose of the comparison with previous studies, we will also show the analytic expression of $\Delta(0)$ for the general axion models \eqref{GENACa}, which is consistent with previous numerical results~\cite{Wu:2021mkk}.
}

This paper is organized as follows. 
In section 2, we introduce the holographic model with magnetic fields and present the determinant method for the quasi-normal modes. 
In section 3, we study the breakdown of magneto-hydrodynamics with the quasi-normal mode computations and determine ($\omega_{\text{eq}}$, $k_{\text{eq}}$). We also show how to obtain $\omega_{\text{eq}}$ analytically from the infra-red Green's function. 
In section 4 we compare our results with the previous study of the breakdown of energy diffusive hydrodynamics~\cite{Arean:2020eus}. 
Section 5 is devoted to conclusions.

%%%%%%%%%%%%%%%%%%%%%%%%%%%%%%%%%%%%%%
%    
%%%%%%%%%%%%%%%%%%%%%%%%%%%%%%%%%%%%%%
\section{Holographic setup}

\subsection{Model}

We consider the Einstein-Maxwell system in (3+1) dimensions
\begin{equation}\label{ACTIONH}
\begin{split}
S = \int \dd^4x \sqrt{-g} \,\left( R \,+\, 6 \,-\, \frac{1}{4} F^2 \right) \,,
\end{split}
\end{equation}
with a background metric
\begin{equation}\label{BGMET}
\begin{split}
\dd s^2 =  -f(r)\, \dd t^2 +  \frac{1}{f(r)} \, \dd r^2  + r^2 (\dd x^2 + \dd y^2) \,,\quad   A = -\frac{H}{2} y \,\dd x \,+\, \frac{H}{2} x \, \dd y \,,
\end{split}
\end{equation}
where $H$ is the external magnetic fields that break translational invariance.
The blackening factor $f(r)$ is 
\begin{equation}\label{BCF}
\begin{split}
 f(r)\,= r^2 - \frac{m_{0}}{r} \,+ \, \frac{H^2}{4\,r^2} \,, \quad m_{0} = r_{h}^3\left( 1 +  \frac{H^2}{4\, r_{h}^4} \right) \,,
\end{split}
\end{equation}
where $m_{0}$ is determined by the condition $f(r_{h})=0$.

Thermodynamic quantities~\cite{Kim:2015wba} including the Hawking temperature $T$ read
\begin{align}\label{HAWKINGT}
 T \,=\, \frac{1}{4\pi} \left( 3\,r_{h} \,-\, \frac{H^2}{4\,r_{h}^3}  \right) \,, \quad \epsilon \,=\, 2 \,r_{h}^3  + \frac{H^2}{2\,r_{h}}   \,, \quad P \,= \, r_{h}^3 - \frac{3 H^2}{4 \, r_{h}} \,, \quad s \,=\, 4\pi r_{h}^2   \,,
\end{align}
where $(\epsilon, P, s)$ are the energy density, pressure, and entropy density respectively.
Note that these quantities satisfy the following Smarr-like relation as
\begin{align}\label{SMARR}
 \epsilon + P = s \, T \,.
\end{align}
%

%
%For later use, we also introduce the first-order transport coefficient $\sigma_{Q}$ given by the fluid/gravity correspondence~\cite{Blake:2015hxa}
% 
%\begin{align}\label{EQSET3}
%\begin{split}
%\sigma_{Q} = \left(\frac{s T}{\epsilon + P}\right)^2 = 1 \,,
%\end{split}
%\end{align} 
%
%where we used the Smarr relation \eqref{SMARR} in the second equality.

\subsection{Hydrodynamic mode}

We can study the holographic dual of hydrodynamic modes of \eqref{ACTIONH}  with the fluctuations 
\begin{align}\label{}
\begin{split}
g_{\mu\nu} \,\rightarrow\, g_{\mu\nu} + \delta g_{\mu\nu} \,, \quad A_{\mu} \,\rightarrow\, A_{\mu} + \delta A_{\mu} \,.
\end{split}
\end{align} 
At the linearized fluctuation level of the Einstein equations and the Maxwell equation, one can find two sets of decoupled fluctuations:
\begin{align}\label{CHNNELCLASSIFICATION}
\begin{split}
&\text{(Sound channel):} \quad\,\,\,\,  \{\delta g_{tt}, \,\delta g_{tx}, \,\delta g_{xx}, \, \delta g_{yy},\, \delta A_{y}\} \,, \\
&\text{(Shear channel):} \qquad  \{\delta g_{ty}, \, \delta g_{xy}, \,\delta A_{t}, \, \delta A_{x}\} \,. \\
\end{split}
\end{align} 
In this paper, we focus on the sound channel related to the energy diffusion mode with the wave vector in the $x$-direction:
\begin{align}\label{FLUCOURSETUP}
\begin{split}
\delta g_{tt} &= h_{tt}(r) \,e^{-i \, \omega \, t + i \, k \, x} \,,\, \quad  \delta g_{tx} = h_{tx}(r) \,e^{-i \, \omega \, t + i \, k \, x} \,,  \quad
\delta g_{xx} = h_{xx}(r) \,e^{-i \, \omega \, t + i \, k \, x} \,, \\
\delta g_{yy} &= h_{yy}(r) \,e^{-i \, \omega \, t + i \, k \, x} \,, \quad\delta A_{y} = a_{y}(r) \,e^{-i \, \omega \, t + i \, k \, x} \,.
\end{split}
\end{align} 
%

%\paragraph{Hydrodynamic mode at finite magnetic fields:}

At finite magnetic fields, \eqref{ACTIONH} produces the quasi-normal modes well matched with the magneto-hydrodynamics. We refer the readers to \cite{Jeong:2021zhz} for further details on this. See appendix \ref{appendixa} for the brief review of the magneto-hydrodynamics. 

The hydrodynamic mode of the sound channel \eqref{FLUCOURSETUP} corresponds to the energy diffusive mode at the small $k$ limit as
\begin{align}\label{SD2}
\omega = -i \,D \, k^2 \,.
\end{align} 
The energy diffusion constant $D$ is given \cite{Blake:2015hxa,Li:2019bgc,Jeong:2021zhz} as
\begin{align}\label{}
\begin{split}
D &:= \frac{\kappa}{c_{\rho}} \,, \quad \kappa = \frac{s^2 \, T}{H^2} \,, \quad c_{\rho} := T\frac{\partial s}{\partial T}  \,, 
\end{split}
\end{align} 
where $\kappa$ is the thermal conductivity and $c_{\rho}$ is the specific heat.
Moreover, using
\begin{align}\label{}
\begin{split}
\frac{\partial s}{\partial T} \,=\, (8\pi r_{h}) \frac{\partial r_{h}}{\partial T}   \,,  \quad \frac{\partial r_{h}}{\partial T} \,=\, \frac{16\pi r_{h}^4}{12 r_{h}^4 + 3 H^2} \,,
\end{split}
\end{align} 
we can express the energy diffusion constant\footnote{This diffusion constant becomes the same as $\frac{\partial P}{\partial \epsilon} \frac{\epsilon + P}{\sigma_{Q}H^2}$ in the small $H/T^2$ limit, where the first-order transport coefficient $\sigma_{Q}=1$ in small $H/T^2$ limit. See \cite{Jeong:2021zhz}.} as
\begin{align}\label{DCF}
\begin{split}
D \,=\,  \frac{3 r_{h}^3}{2 H^2} + \frac{3}{8 r_{h}}  \,.
\end{split}
\end{align} 
%

% 
%\begin{align}\label{DCF}
%D \,=\, \frac{\partial P}{\partial \epsilon} \frac{\epsilon + P}{\sigma_{Q}H^2} \,=\, \frac{3}{8\,r_{h}} \left( 1 + \frac{4 r_{h}^4}{H^2} \right) \,,
%\end{align} 
%
%where \eqref{HAWKINGT}-\eqref{EQSET3} is used in the second equality\footnote{This diffusion constant is the same as $D=\kappa/c_{\rho}$ where $\kappa$ is the thermal conductivity and $c_{\rho}$ is the specific heat. See \cite{Jeong:2021zhz}.}. 

%%%%%%%%%%%%%%%%%%%%%%%%%%%%%%%%%%%%%%
%    
%%%%%%%%%%%%%%%%%%%%%%%%%%%%%%%%%%%%%%
\subsection{The determinant method}
The determinant method for the magnetically charged black hole is given in \cite{Jeong:2021zhz}.
For the convenience of the reader, we also present it as follows.

\paragraph{Gauge-invariant perturbation:}
In order to study the hydrodynamic mode \eqref{SD2} in holography, we choose the following deffeomorphism and gauge-invariant combinations~\cite{Buchbinder:2008dc,Buchbinder:2009aa}:
\begin{align}
\begin{split}
Z_{H} &:= \frac{4 k}{\omega} \, h_{t}^{x} \,+\,  2 h_{x}^{x} - \left( 2 - \frac{k^2}{\omega^2}\frac{f'(r)}{r} \right) h_{y}^{y}  + \frac{2k^2}{\omega^2}\frac{f(r)}{r^2} h_{t}^{t}  \,, \\
Z_{A} &:= a_{y}  \,+\, \frac{i H}{2k} \left(h_{x}^{x}-h_{y}^{y}\right) \,,
\end{split}
\end{align}
where we raised an index on fluctuation fields using the background metric \eqref{BGMET}.
Then we can obtain the gauge invariant second order equations for $Z_{H}$ and $Z_{A}$ of the following form:
\begin{align}\label{ZAZHEOM}
\begin{split}
&0 \,=\, A_{H}\,Z_{H}'' \,+\, B_{H}\,Z_{H}' \,+\, C_{H}\,Z_{H} \,+\, D_{H}\,Z_{A}' \,+\, E_{H}\,Z_{A} \,, \\
&0 \,=\, A_{A}\,Z_{A}'' \,\,+\, B_{A}\,Z_{A}' \,\,+\, C_{A}\,Z_{A} \,\,+\, D_{A}\,Z_{H}' \,\,+\, E_{A}\,Z_{H} \,.
\end{split}
\end{align}
Since the coefficients of equations are lengthy and cumbersome we will not write them in the paper.

Next, we solve the equations of motion \eqref{ZAZHEOM} with two boundary conditions: one from incoming boundary condition at the horizon and the other from the AdS boundary.
First, near the horizon ($r\rightarrow r_{h}$) the variables are expanded as 
\begin{align}\label{APPENHORIZON}
\begin{split}
Z_{H} = (r-r_{h})^{\nu_{\pm}} \left( Z_{H}^{(I)} \,+\, Z_{H}^{(II)} (r-r_{h}) \,+\, \dots   \right ) \,, \\
Z_{A} = (r-r_{h})^{\nu_{\pm}} \left( Z_{A}^{(I)} \,+\, Z_{A}^{(II)} (r-r_{h}) \,+\, \dots   \right ) \,,
\end{split}
\end{align}
where $\nu_{\pm}:= \pm i\omega/4 \pi T$ and we choose $\nu_{-}:= - i\omega/4 \pi T$ for the incoming boundary condition.
After plugging \eqref{APPENHORIZON} into equations \eqref{ZAZHEOM}, one can find that higher order horizon coefficients are determined by two independent horizon variables : $(Z_{H}^{(I)}, Z_{A}^{(I)})$.

Near the AdS boundary ($r\rightarrow \infty$), the asymptotic behavior of solutions read
\begin{align}\label{}
\begin{split}
&Z_{H} = Z_{H}^{(S)} \, r^{0} \,(1 \,+\, \dots) \,+\, Z_{H}^{(R)} \, r^{-3} \,(1 \,+\, \dots) \,, \\
&Z_{A} = Z_{A}^{(S)} \, r^{0} \,(1 \,+\, \dots) \,+\, Z_{A}^{(R)}\, r^{-1} \,(1 \,+\, \dots) \,,
\end{split}
\end{align}
where the superscript $(S)$ refers to a source term while $(R)$ denotes a response term.

\paragraph{The determinant method.}
We can compute the quasi-normal modes by employing {the determinant method~\cite{Kaminski:2009dh}}. Using the shooting method, we can construct the matrix of sources with the source terms near the boundary:
\begin{align}\label{APPENSMATA}
\begin{split}
S = \left(\begin{array}{cc}Z_{H}^{(S)(I)} & Z_{H}^{(S)(II)} \\Z_{A}^{(S)(I)} & Z_{A}^{(S)(II)}\end{array}\right) \,,
\end{split}
\end{align}
where the $S$-matrix is $2\times2$ matrix because we can get two independent solutions with two independent shooting variables at the horizon \eqref{APPENHORIZON}. Note that $I (II)$ in \eqref{APPENSMATA} denotes that the source terms are obtained by the $I (II)$ th shooting.

Finally, the quasi-normal mode spectrum, i.e. the dispersion relation for which the holographic Green's functions have a pole, can be given by the value of ($\omega, k$) for which the determinant of $S$-matrix \eqref{APPENSMATA} vanishes.

\section{Breakdown of magneto-hydrodynamics}

%%%%%%%%%%%%%%%%%%%%%%%%%%%%%%%%%%%%%%
%    
%%%%%%%%%%%%%%%%%%%%%%%%%%%%%%%%%%%%%%
\subsection{Infra-red modes}

%\subsection{Near-horizon perturbation equations}

In this section, we study the perturbation equation \eqref{ZAZHEOM} in the $\text{AdS}_2 \times \text{R}^2$ spacetime that emerges near the horizon at low temperatures.
We will closely follow the method given in the appendix A in \cite{Arean:2020eus}.

%\paragraph{The purpose of this section:}
By solving the perturbation equation in the extremal geometry (near the horizon at low temperature), we will obtain the infra-red Green's function $\mathcal{G}_{IR}$ analytically with an operator of dimension $\Delta(k)$ at the infra-red fixed point of the field theory. 
From this analysis, we will show the analytic equilibration frequency $\omega_{\text{eq}}$ at low temperature is
\begin{align}\label{}
\begin{split}
 \omega_{\text{eq}} \,=\, -i \, 2 \pi T\, \Delta(k=0) \,.
\end{split}
\end{align}

\paragraph{The extremal geometry.}
Let us first discuss the small temperature condition. From \eqref{HAWKINGT}, one can check the temperature becomes zero ($T=0$) at 
\begin{align}\label{}
\begin{split}
r_{h} = r_{e} \,,\quad  r_{e} := \left(\frac{H}{\sqrt{12}}\right)^{1/2}\,.
\end{split}
\end{align}
Thus, by putting the following relation into the Hawking temperature in \eqref{HAWKINGT}
\begin{align}\label{}
\begin{split}
 T \,=\, 0 \,+\, \epsilon \, \delta T \,, \quad r_{h} \,=\, r_{e} \,+\, \epsilon \, \zeta_{h} \,,
\end{split}
\end{align}
and take $\epsilon\rightarrow0$, we can find the small temperature correction $\zeta_{h}$ as
\begin{align}\label{}
\begin{split}
\zeta_{h} \,=\, \frac{\pi}{3} \delta T \,.
\end{split}
\end{align}

Next, in order to obtain the extremal geometry, one can consider the following coordinate transformation \cite{Arean:2020eus,Faulkner:2009wj} in \eqref{BGMET}
\begin{align}\label{COT}
\begin{split}
 r \,=\, r_{e} \,+\, \epsilon \, \zeta \,, \quad r_{h} \,=\, r_{e} \,+\, \epsilon \, \zeta_{h} \,,\quad t \,=\, \frac{u}{\epsilon} \,,
\end{split}
\end{align}
and take $\epsilon\rightarrow0$. Then we can change the coordinate from ($t, r$) to ($u, \zeta$) as follows
\begin{equation}\label{EXTG}
\begin{split}
\dd s^2 =  -\frac{\zeta^2}{L^2} \left( 1-\frac{\zeta_{h}}{\zeta} \right)^2 \, \dd u^2 +  \frac{L^2}{\zeta^2 \, \left( 1-\frac{\zeta_{h}}{\zeta} \right)^2} \, \dd \zeta^2  + r_{e}^2 (\dd x^2 + \dd y^2) \,,
\end{split}
\end{equation}
where the AdS$_{2}$ radius of curvature is $L^2=1/6$. Note that this geometry corresponds to the $\text{AdS}_2 \times \text{R}^2$ geometry with the small temperature correction ($\zeta_{h}$), and $\zeta$ runs from $\zeta = \zeta_{h}$ to $\zeta = \infty$ (the $\text{AdS}_{2}$ boundary).

\paragraph{The perturbation equation in the extremal geometry.}

Using the following coordinate transformation,
\begin{align}\label{COT2}
\begin{split}
 r \,=\, r_{e} \,+\, \epsilon \, \zeta \,, \quad r_{h} \,=\, r_{e} \,+\, \epsilon \, \zeta_{h} \,,\quad \omega \,=\, \epsilon \, \zeta_{\omega}  \,,
\end{split}
\end{align}
we can obtain the perturbation equations \eqref{ZAZHEOM} in the extremal geometry \eqref{EXTG} in the $\epsilon\rightarrow0$ limit.
Note that \eqref{COT2} is the Fourier transformed version of \eqref{COT}.

At the leading order in $\epsilon$, one can check only the second perturbation equation in \eqref{ZAZHEOM} survives: the first perturbation equation starts at $\mathcal{O}(\epsilon)$ order.
In particular, the survived equation of motion is composed of only one field $Z_{A}(\zeta)$:
\begin{align} \label{prerq1}
\qquad &\partial_{\zeta}^2 Z_{A} + \left( \frac{2\zeta}{\zeta^2 - \zeta_{h}^2} \right) \partial_{\zeta} Z_{A} + \left(  \frac{\zeta_{\omega}^2}{36\,(\zeta^2 - \zeta_{h}^2)^2} - \frac{k^2}{\sqrt{3}\,H \,(\zeta^2 - \zeta_{h}^2)} \right) Z_{A} = 0 \,.
\end{align}

Near the AdS$_{2}$ boundary ($\zeta \rightarrow \infty$), the solution of the equation is expanded as 
\begin{align}\label{bdasd3}
\begin{split}
Z_{A} \,=\, Z^{(S)} \, \zeta^{\Delta(k)-1}  + Z^{(R)} \, \zeta^{-\Delta(k)}  \,,
\end{split}
\end{align}
where an operator of dimension $\Delta(k)$ at the infra-red fixed point of the field theory is
\begin{align}\label{LATdd}
\begin{split}
\Delta(k) = \frac{1}{2} \left( 1 + \sqrt{1 + \frac{4\,k^2}{\sqrt{3}\,H}} \right) \,.
\end{split}
\end{align}
%

%\subsection{The infra-red modes}

\paragraph{The infra-red Green's function:}
The infra-red Green's function can be found explicitly by solving the equations \eqref{prerq1} and imposing the usual AdS/CFT rules\footnote{The general solution has the associated Legendre function of the second kind which corresponds to the out-going solution at $\zeta = \zeta_{h}$, so we should discard it.} at the AdS$_{2}$ boundary ($\zeta \rightarrow \infty$) \cite{Faulkner:2009wj,Hartnoll:2012rj} 
\begin{align}
\begin{split}
\mathcal{G}_{IR} \propto \frac{Z^{(R)}}{Z^{(S)}} \,,
\end{split}
\end{align}
where  $Z^{(S)}$, $Z^{(R)}$ are coefficients in \eqref{bdasd3}.
Let us show the explicit form of the infra-red Green's function 
\begin{align}\label{IRG}
\begin{split}
\mathcal{G}_{IR} =   \frac{2\Delta(k)-1}{2} \left(\frac{6}{\pi}\right)^{1-2\Delta(k)}    \delta T^{2 \, \Delta (k) -1}  \frac{ \Gamma \left(  \frac{1}{2} - \Delta(k) \right) \Gamma \left(  \Delta(k)  - \frac{i\zeta_{\omega}}{2\pi  \delta T}\right) }{ \Gamma \left(  \frac{1}{2} + \Delta(k) \right) \Gamma \left( 1- \Delta(k)  - \frac{i\zeta_{\omega}}{2\pi  \delta T}\right) }\,.
\end{split}
\end{align}
%
%where we replace notations as ($\delta T \rightarrow T$, $\zeta_{\omega} \rightarrow \omega$). 
%In the last section, we compare \eqref{LATdd}-\eqref{IRG} with the axion model case~\cite{Arean:2020eus}.

From \eqref{IRG}, the infra-red Green's function, $\mathcal{G}_{IR}$, gives poles along the imaginary frequency axis at the locations
\begin{align}\label{IRPOLE}
\begin{split}
\zeta_{\omega, \, n} \,=\, -i \, 2\pi\, \delta T \left( n \,+\, \Delta (k) \right)\,, \qquad n \,=\, 0,1,2,\dots \,.
\end{split}
\end{align}
In order to reconstruct the full retarded Green's function ($G^{R}$) analytically, giving the quasi-normal mode beyond the small temperature, one must extend the near-horizon solutions through the rest of the spacetime, which is difficult to do. 
However, for the purpose of studying the low-temperature behavior, it might be enough to observe that $G^R$ exhibits poles from \eqref{APPENSMATA}, where its locations approach those of the poles of $\mathcal{G}_{IR}$ \eqref{IRPOLE} as $(k \rightarrow 0,\, T \rightarrow 0)$~\cite{Arean:2020eus}\footnote{Dividing \eqref{SD2} with $T$, we have $\omega/T = -i D \, k^2/T$. Then, the relevant low $T$ limit is $(k \rightarrow 0,\, T \rightarrow 0)$ to keep $k^2/T$ fixed and therefore $\omega/T$ fixed. See more details in the appendix A of \cite{Arean:2020eus}.}. 
In other words, this means that in this limit $G^{R}$ exhibits poles at 
\begin{align}\label{irqnms1}
\begin{split}
\omega_{n} \,=\, -i \, 2\pi\, T \left( n \,+\, \Delta (0) \right)\,, \qquad n \,=\, 0,1,2,\dots \,,
\end{split}
\end{align}
where $( \omega_{n}, \, T)$ may approach to $( \zeta_{\omega, \, n} \,,\, \delta T)$ with $\Delta(0)=1$ \eqref{LATdd}.
In the next section, we will show this analytic result is consistent with the numerical results of the quasi-normal mode spectrum.

%%%%%%%%%%%%%%%%%%%%%%%%%%%%%%%%%%%%%%
%    
%%%%%%%%%%%%%%%%%%%%%%%%%%%%%%%%%%%%%%
\subsection{Non-hydrodynamic poles and the collision}
\paragraph{Non-hydrodynamic modes.}
In order to study the breakdown of magneto-hydrodynamics, we need to identify the non-hydrodynamic modes whose longest-lived mode (the first non-hydro mode) will make the collision with the hydrodynamic mode. 
%So it might be instructive to see the origin of the non-hydrodynamic modes. 

In particular, as in the hydrodynamic mode \eqref{SD2}, it is tempting to obtain the expression of the non-hydrodynamic mode in the small wave vector regime.
As long as we consider the low temperature regime, it has been discovered the non-hydrodynamic modes follow \eqref{irqnms1} at small wave vector~\cite{Arean:2020eus,Wu:2021mkk}, i.e., the origin of the non-hydrodynamic mode at low $T$ is associated to the AdS$_{2}$ geometry.

\begin{figure}[]
\centering
     {\includegraphics[width=7.2cm]{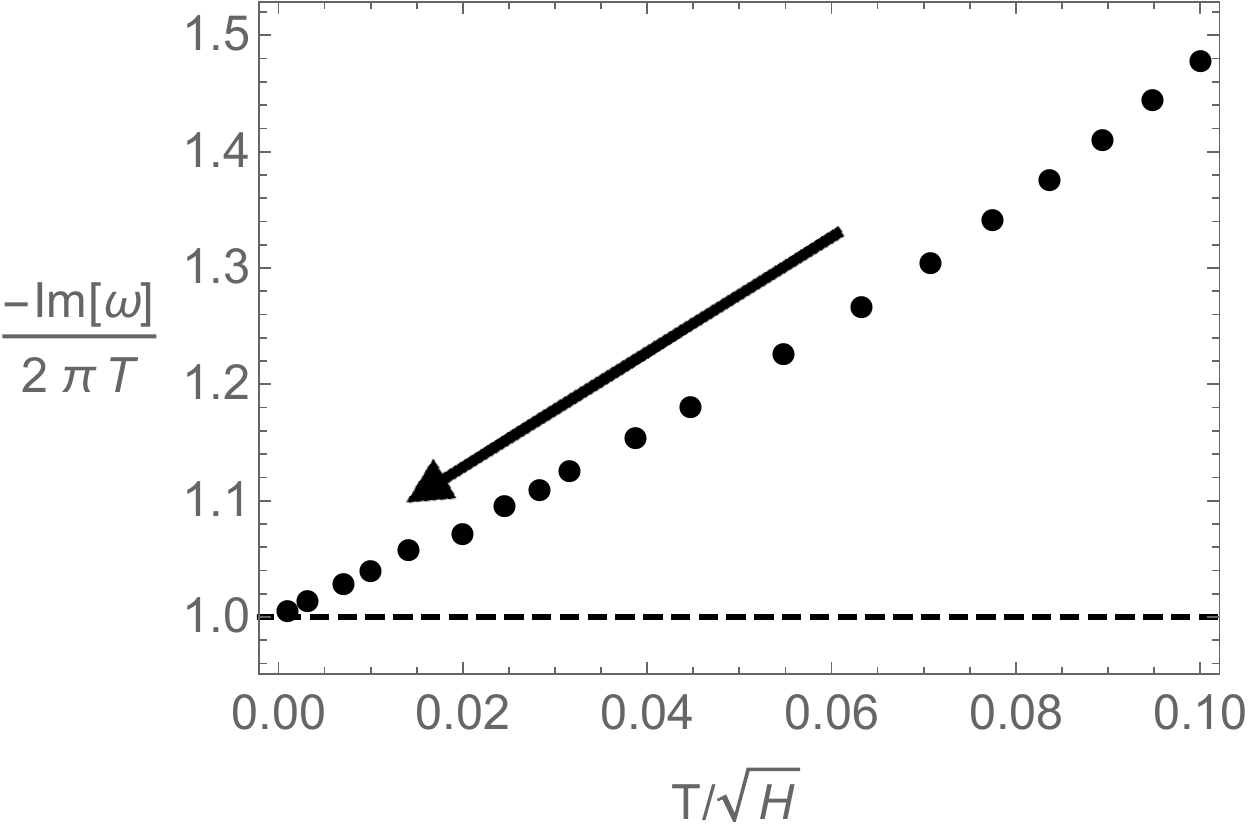} \label{}}
 \caption{The first non-hydrodynamic mode at $k/T=1/10$. Black dots are quasi-normal modes and the dashed line is $\omega_{0}$ \eqref{irqnms1}.}\label{STSKF}
\end{figure}

We find this is also valid in the presence of magnetic fields.
In Fig. \ref{STSKF}, we display the first non-hydro mode\footnote{The higher non-hydro mode is also well matched with $\omega_{n}$ \eqref{irqnms1}.} in agreement with $\omega_{0}$ \eqref{irqnms1} at $T/\sqrt{H}\rightarrow0$, which is responsible for the breakdown of hydrodynamic.
%: so the breakdown will be associated to the AdS$_{2}$ geometry even in the presence of magnetic fields.
%
\begin{figure}[]
\centering
     \subfigure[Quasi-normal mode]
     {\includegraphics[width=6.6cm]{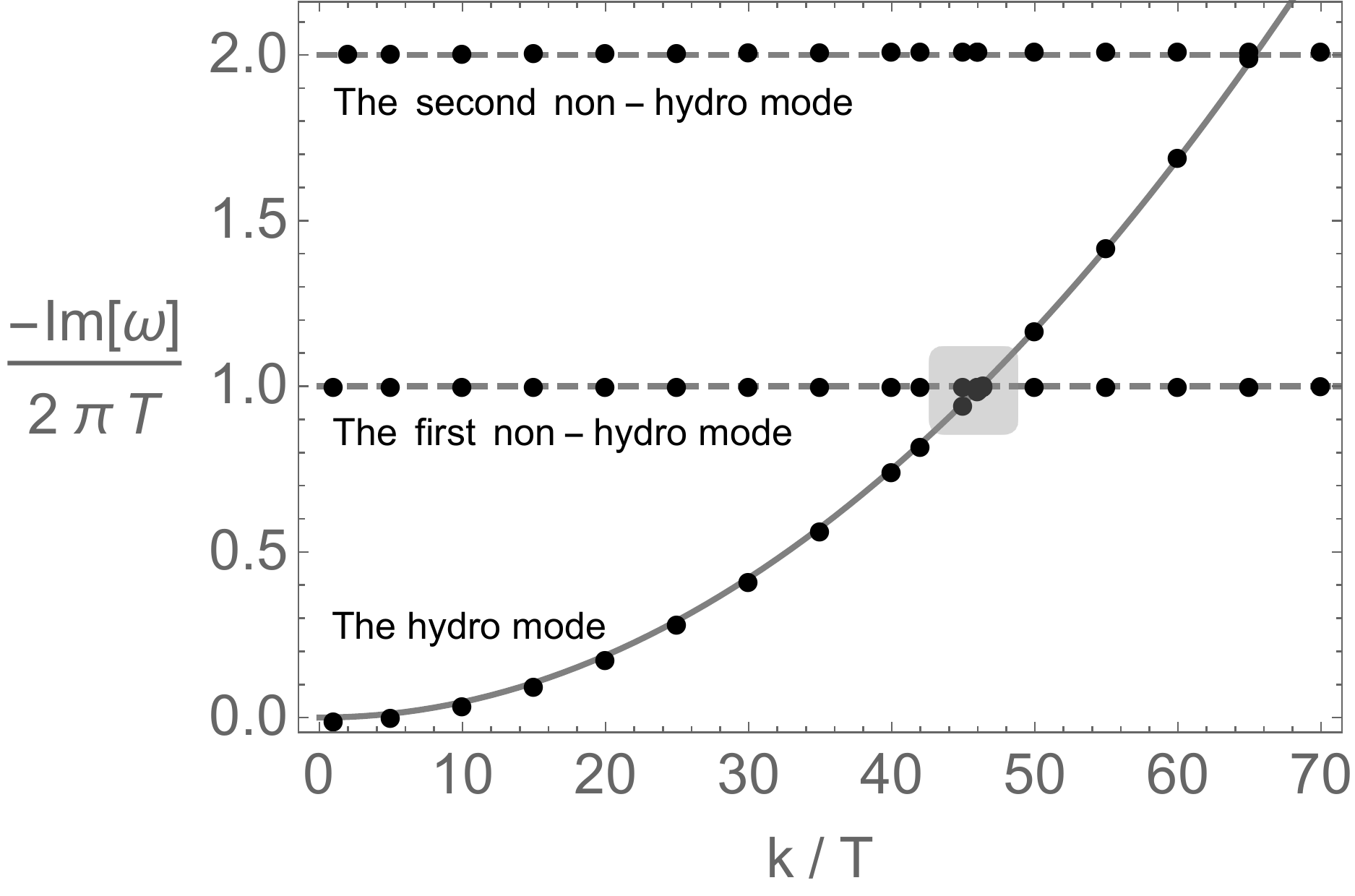} \label{STSKF2}}
          \subfigure[A zoom of the gray square region]
     {\includegraphics[width=7.6cm]{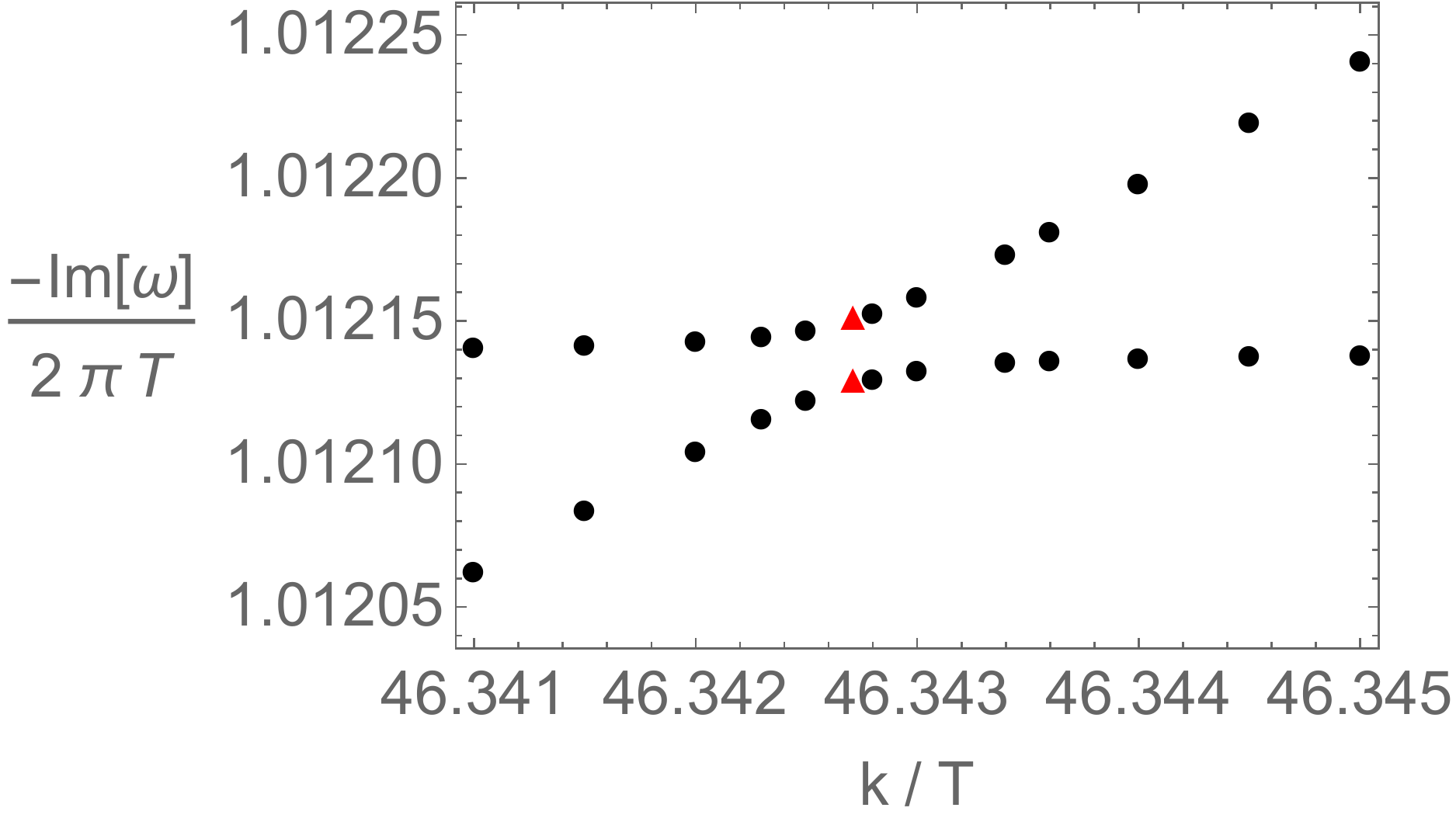} \label{STSKF3}}
 \caption{Quasi-normal modes (black dots) at $H/T^2=10^{5}$. \textbf{Left:} The solid line is the hydrodynamic mode \eqref{SD2} and the dashed line is the non-hydrodynamic mode \eqref{irqnms1}. \textbf{Right:} A zoomed in figure of the gray square region in Fig. \ref{STSKF2}.}\label{}
\end{figure}

\paragraph{The collision between poles.}
Next, let us examine how this first non-hydro mode collides with the hydrodynamic mode.
In Fig. \ref{STSKF2}, we first show the quasi-normal modes at low $T$.
One can see the quasi-normal modes (black dots) are well matched with i) the hydrodynamic mode (solid line) \eqref{SD2}; ii) the non-hydrodynamic mode (dashed line) \eqref{irqnms1}.

The gray square region around $k/T\sim46$ indicates the location in which the collision between the hydro mode and the first non-hydro mode may appear. As we investigate this gray region carefully, one may see the collision cannot be visible in this real $k/T$ plot. 
We display the zoom of this gray region in Fig. \ref{STSKF3}: the first non-hydro mode does not collide with the hydro mode where the red triangle will be used as the guide to find the collision point in the following.

\begin{figure}[]
\centering
     {\includegraphics[width=7.2cm]{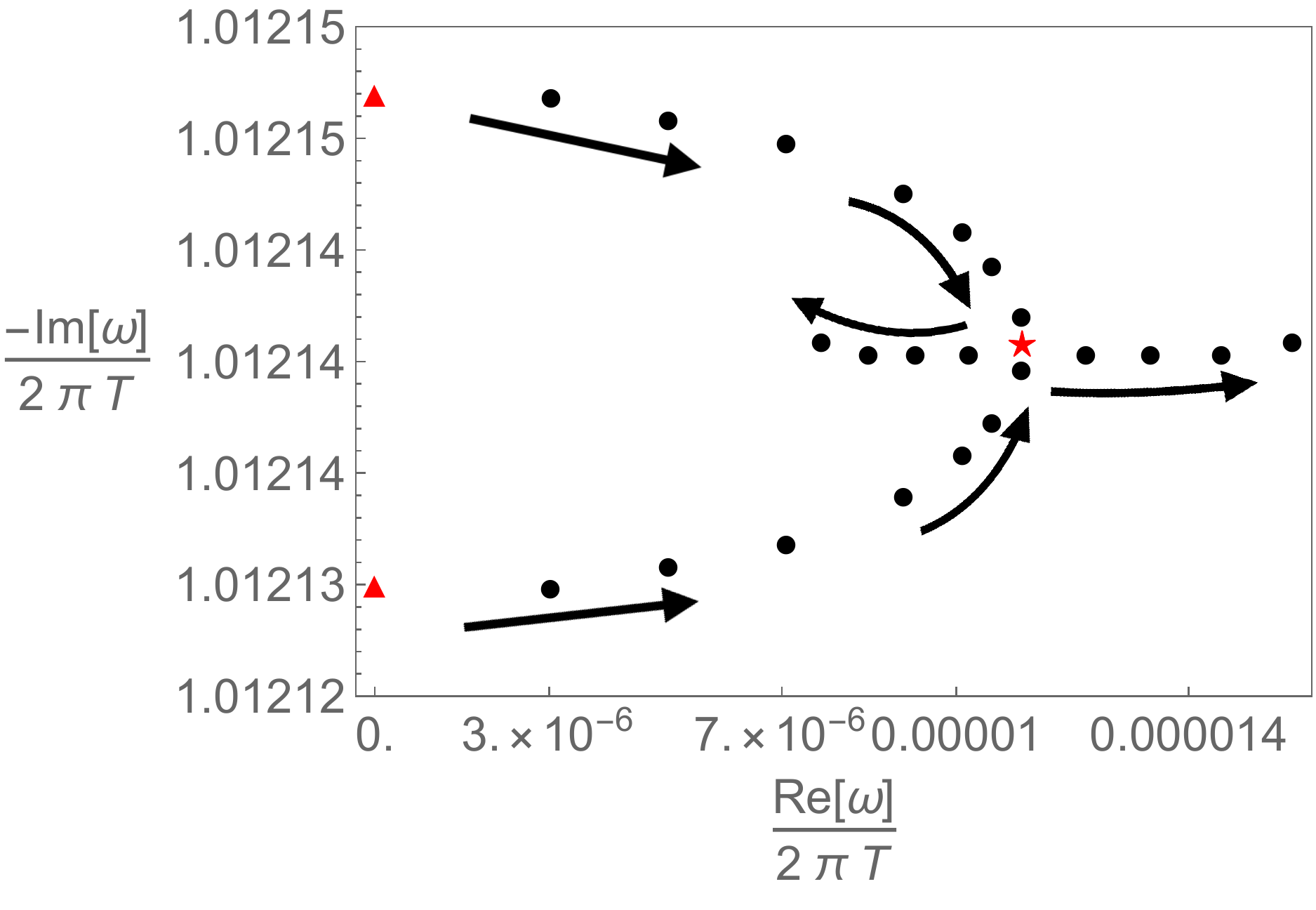} }
 \caption{Quasi-normal modes in the complex $\omega$ plane as $\phi_{k}$ is increased from $0$ (red triangles) to $0.0000116$ at $H/T^2=10^{5}$ and $|k|/T=46.342712$. Red triangles are the same as in Fig. \ref{STSKF3}. Red star indicates the collision point ($\phi_{k}\sim0.00001113$).}\label{Mfig6FIG}
\end{figure}

In order to find out the collision point ($\omega_{\text{c}}, k_{\text{c}}$), we need to extend the quasi-normal mode analysis into the complex plane~\cite{Arean:2020eus}.
Introducing the phase of $k$ ($\phi_{k}$), 
\begin{align}\label{phaseeq}
\begin{split}
k = |k| \,e^{i\,\phi_{k}} \,,
\end{split}
\end{align}
we can examine how the poles move in the complex $\omega$ plane.
In Fig. \ref{Mfig6FIG}, we display the poles at $|k|/T=46.342712$ with changing $\phi_{k}$\footnote{As $\phi_{k}$ approaches to $2\pi$, it is expected that the poles become the same poles for $\phi_{k}=0$.}.
\begin{figure}[]
\centering
     \subfigure[$|k|<|k_c|$]
     {\includegraphics[width=4.8cm]{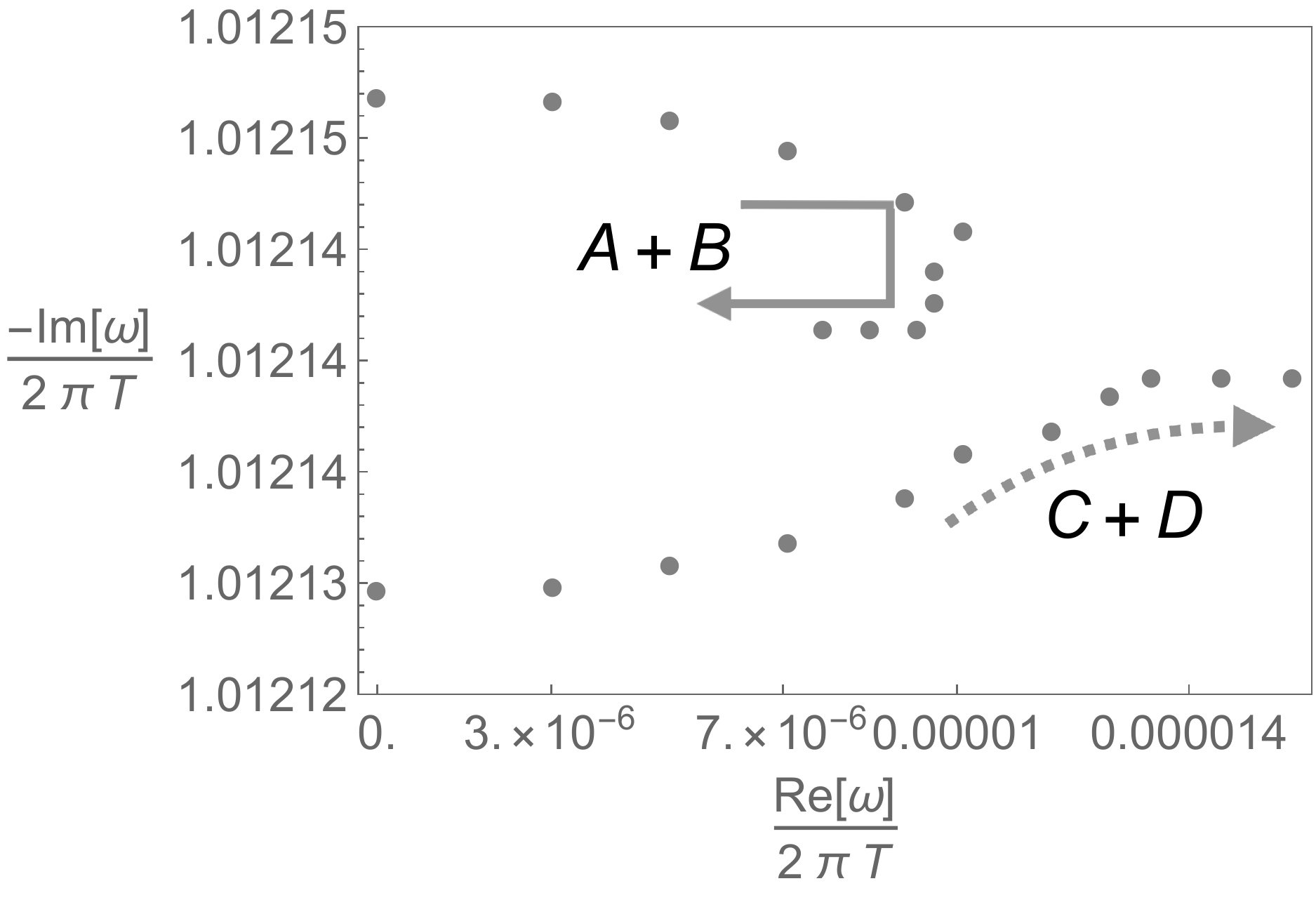} \label{Mfig14FIGa}}
     \subfigure[$|k|=|k_c|$]
     {\includegraphics[width=4.8cm]{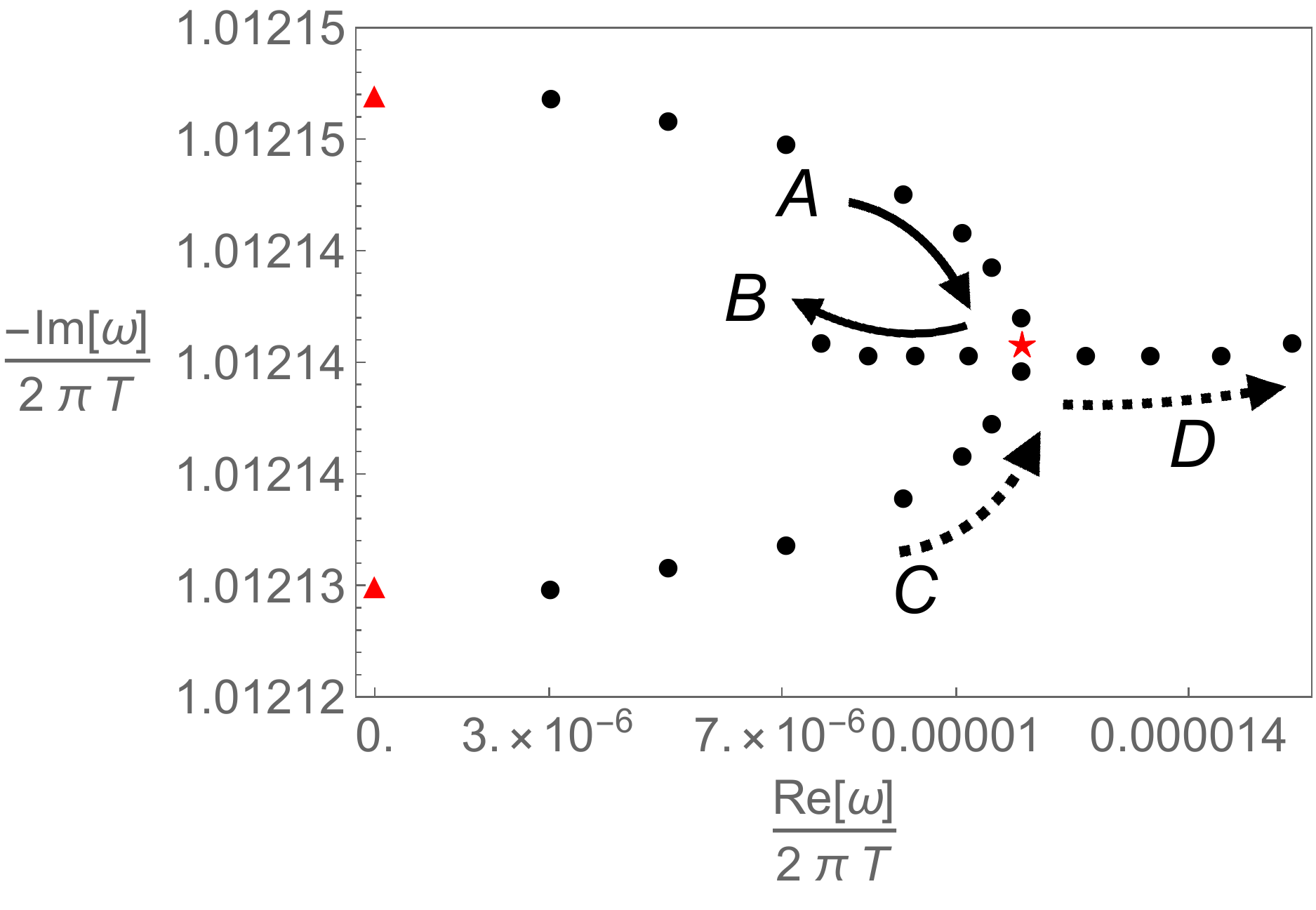} \label{Mfig14FIGb}}
     \subfigure[$|k|>|k_c|$]
     {\includegraphics[width=4.8cm]{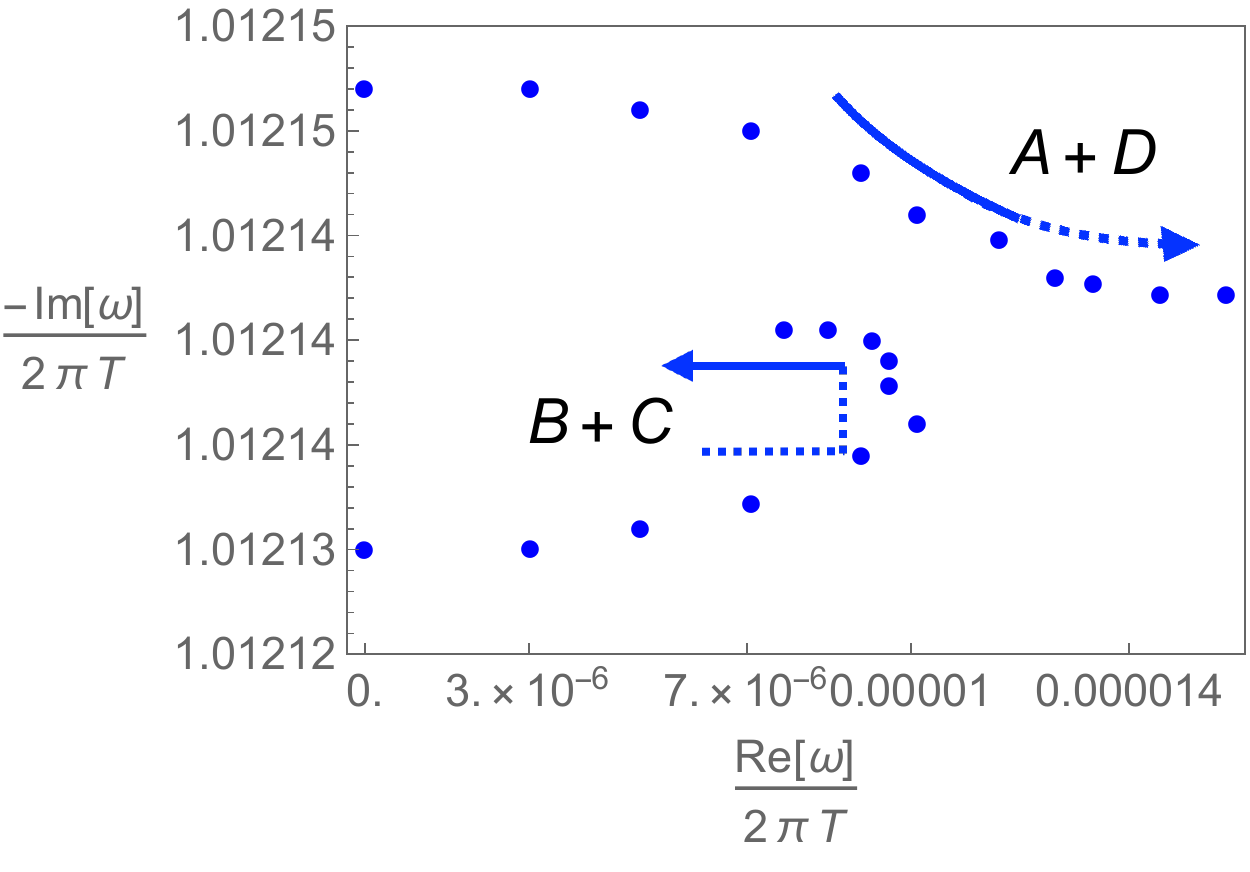} \label{Mfig14FIGc}}
 \caption{Quasi-normal mode near $|k_c|/T$ at $\phi_{k}\in[0, \,0.0000116]$: $|k|-|k_c| = (-1.2\times10^{-5}, \,0, \,1.2\times10^{-5}) \, T$ (left, center, right).}\label{Mfig14FIG}
\end{figure}
As we increase the phase $\phi_{k}$ from $0$ (red triangles) to $0.0000116$, poles make the collision around $\phi_{k}\sim0.00001113$ (red star).
Note that the red triangles ($\phi_{k}=0$) in Fig. \ref{Mfig6FIG} are the same as in Fig. \ref{STSKF3}, so one may imagine putting one additional axis Re$[\omega]$ on Fig. \ref{STSKF3} with \eqref{phaseeq}, and that would be Fig. \ref{Mfig6FIG}.
%think that Fig. \ref{Mfig6FIG} corresponds to the extension of Fig. \ref{STSKF3} into Re$[\omega]$ direction with complex $k$.

The wave number ($|k|/T=46.342712$) used in Fig. \ref{Mfig6FIG} corresponds to $|k_c|/T$ because of the existence of the collision. Let us further investigate the quasi-normal mode near $|k_c|/T$ in Fig. \ref{Mfig14FIG} and summarize the results as
\begin{itemize}
\item{$|k|=|k_c|$: Fig. \ref{Mfig14FIGb}, we label ($A, B, C, D$) for each arrows, which is the same figure as Fig. \ref{Mfig6FIG}. }
\item{$|k|<|k_c|$: Fig. \ref{Mfig14FIGa}, one cannot see the collision and $A$ connects to $B$ while $C$ is with $D$. }
\item{$|k|>|k_c|$: Fig. \ref{Mfig14FIGc}, one cannot also see the collision and $A$ connects to $D$ while $C$ is with $B$. }
\end{itemize}

%%%%%%%%%%%%%%%%%%%%%%%%%%%%%%%%%%%%%%
%    
%%%%%%%%%%%%%%%%%%%%%%%%%%%%%%%%%%%%%%
\subsection{The breakdown of magneto-hydrodynamics}
\begin{figure}[]
\centering
     \subfigure[$\omega_{\text{eq}}$ vs $T$]
     {\includegraphics[width=5.0cm]{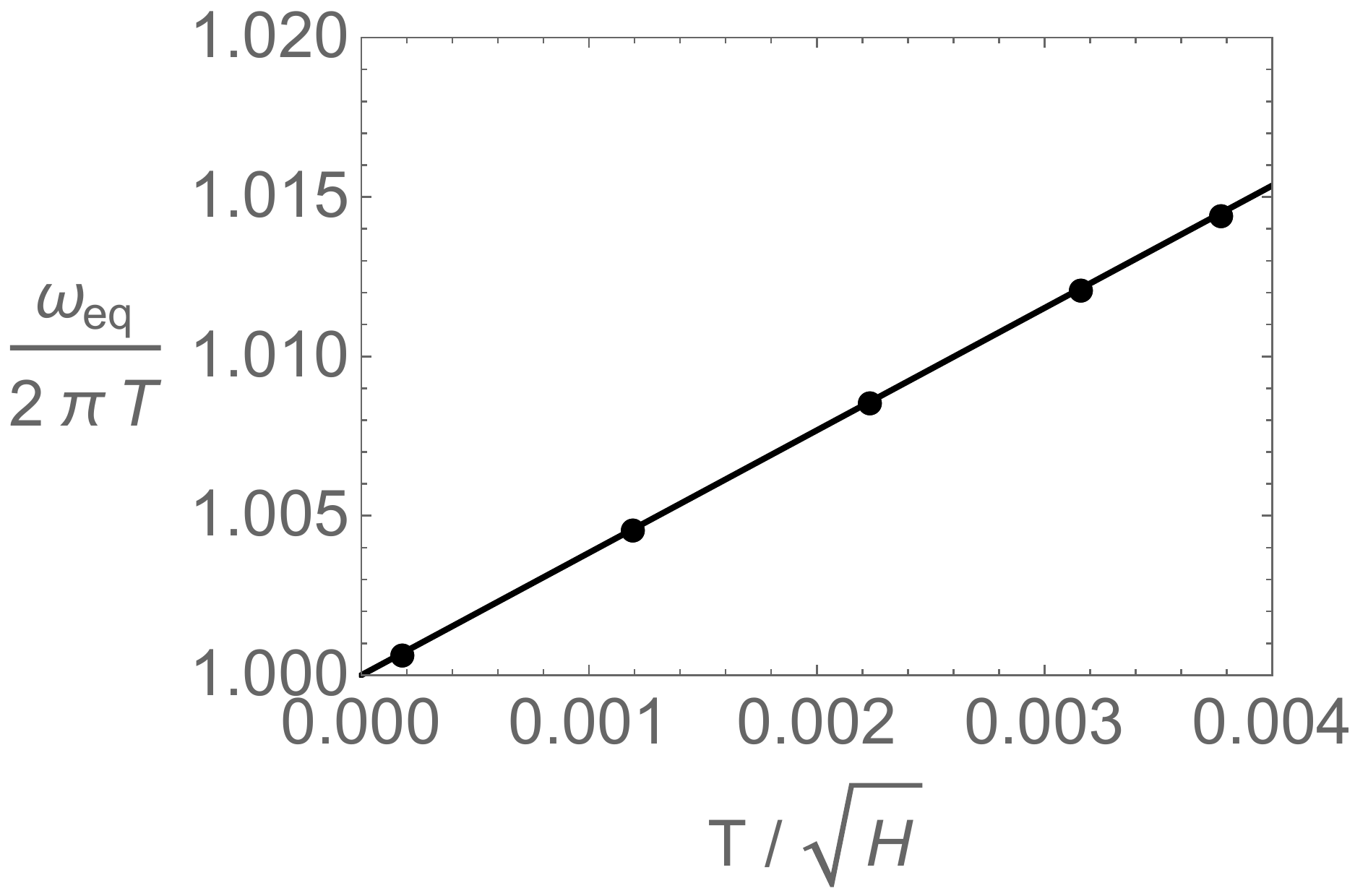} \label{}}
     \subfigure[$k_{\text{eq}}$ vs $T$]
     {\includegraphics[width=4.5cm]{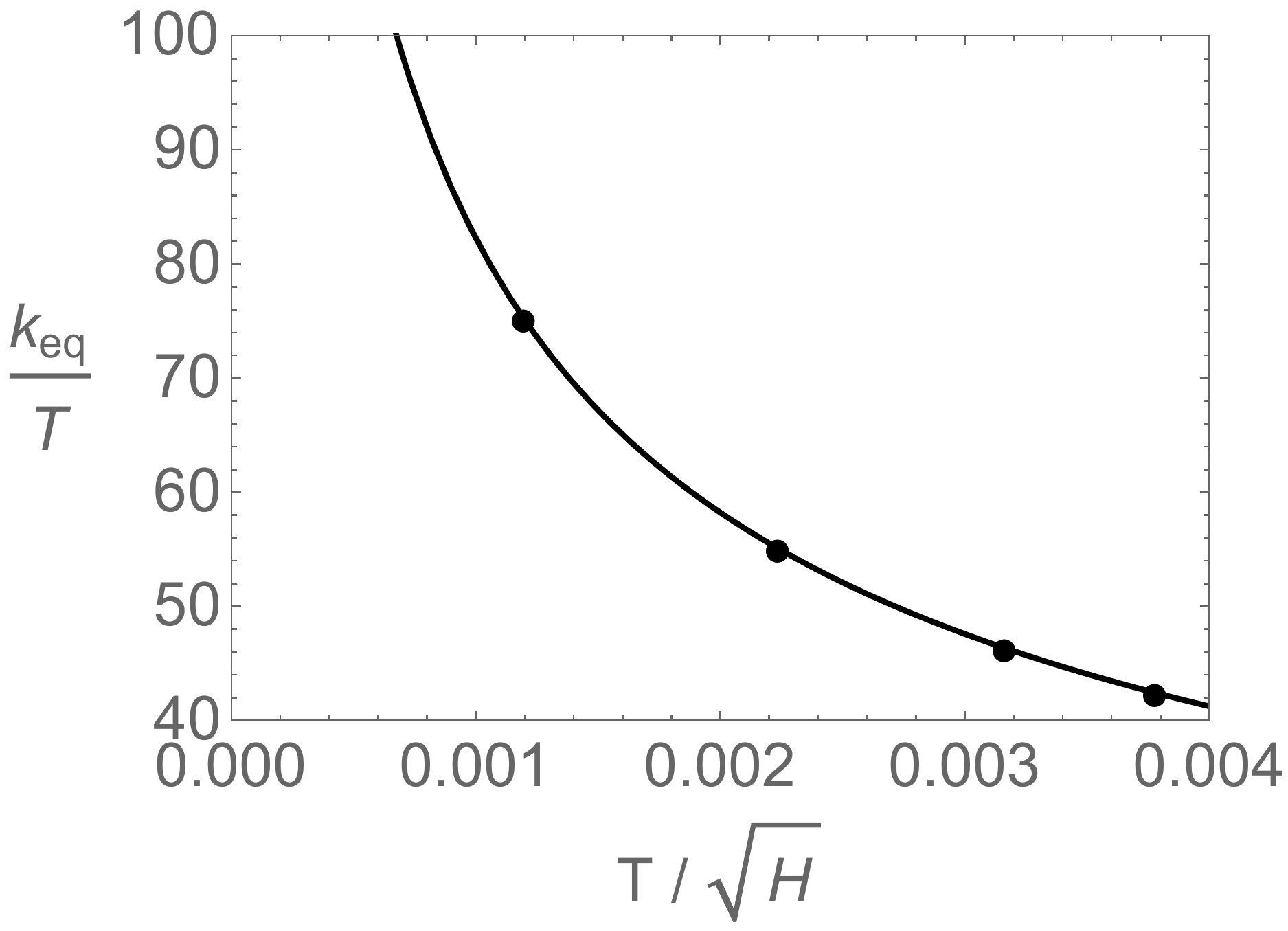} \label{}}
     \subfigure[$\phi_{k}$ vs $T$]
     {\includegraphics[width=4.9cm]{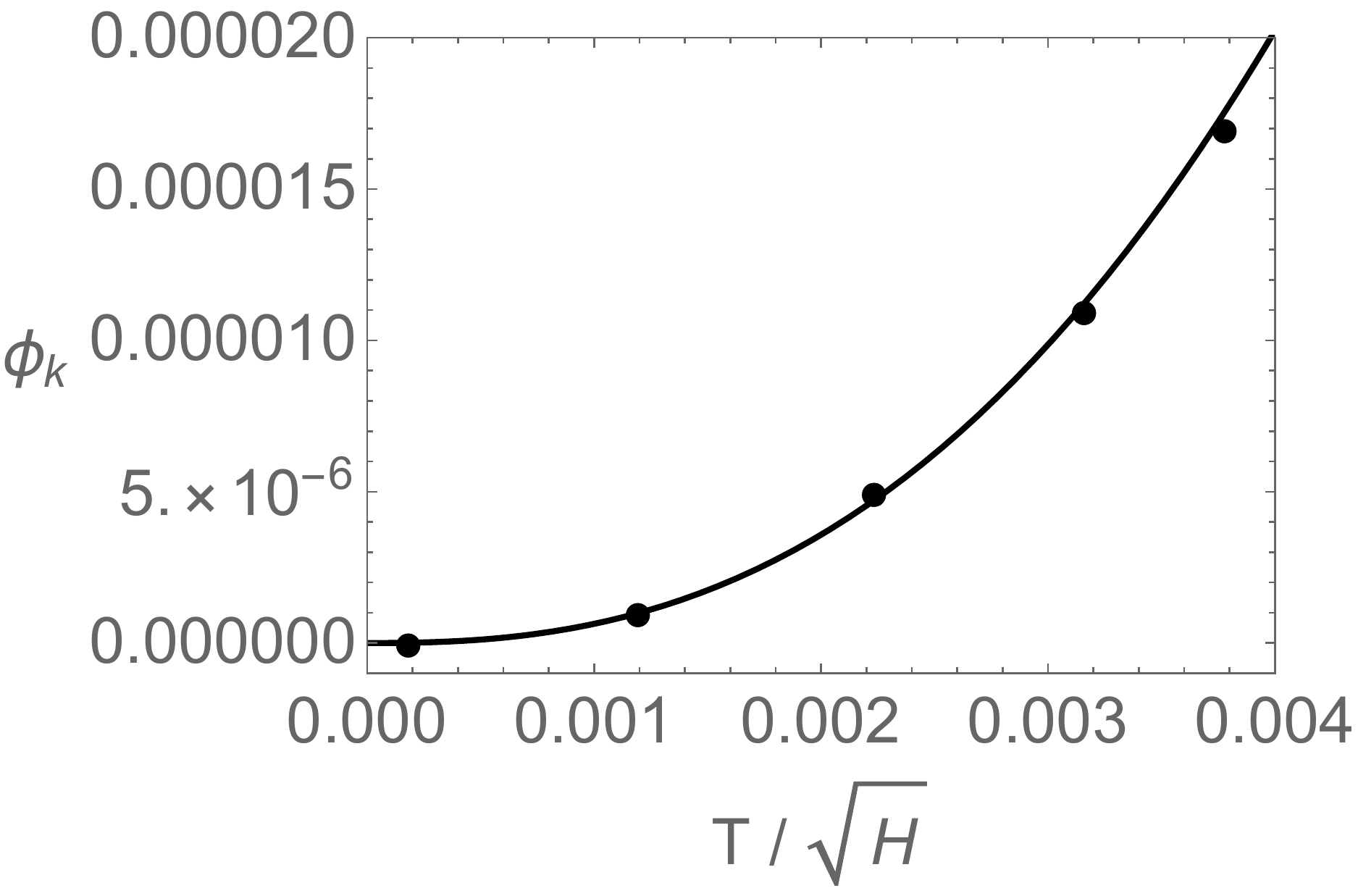} \label{}}
 \caption{The temperature dependence of ($\omega_{\text{eq}},\, k_{\text{eq}}$,\, $\phi_{k}$). Dots are numerical data and black solid lines are fitting curves \eqref{FC1}.}\label{Mfig1111FIG}
\end{figure}

%\paragraph{The breakdown of magneto-hydrodynamics:}
The collision point ($\omega_{\text{c}}, k_{\text{c}}$) signals the breakdown of the hydrodynamics and its absolute value  is defined as ($\omega_{\text{eq}}, k_{\text{eq}}$)~\cite{Arean:2020eus,Wu:2021mkk}
\begin{align}\label{EQFOR}
\begin{split}
\omega_{\text{eq}} := | \omega_{c} | \,,\quad k_{\text{eq}} := | k_{c} |\,.
\end{split}
\end{align}
We show the temperature dependence of the equilibration data ($\omega_{\text{eq}},\, k_{\text{eq}}$,\, $\phi_{k}$) in Fig. \ref{Mfig1111FIG} with the fitting curves
\begin{align}\label{FC1}
\begin{split}
\frac{\omega_{\text{eq}}}{2 \pi T} \,\sim\,  \Delta(0) + \# \, T  \,, \qquad \frac{k_{\text{eq}}}{T} \,\sim\,  \frac{\#}{\sqrt{T}} \,, \qquad \phi_{k} \,\sim\,  \# \, T^{5/2}  \,,
\end{split}
\end{align}
where $\Delta(0)=1$ \eqref{LATdd}\footnote{For the shear diffusion mode~\cite{Arean:2020eus}, it turned out $\Delta(0)$ has the same value as $\Delta(0)=1$. However, the phase $\phi_{k}$ has a different scaling behavior as $\phi_{k} \,\sim\, T^{1/2}$. So the phase seems to depend on the type of the diffusive mode.} and as pointed out in \cite{Arean:2020eus}, the collision point seems to follow 
\begin{align}\label{}
\begin{split}
\omega_{c} =  \omega_{\text{eq}}\, e^{i (\phi_{k} - \frac{\pi}{2})}  \,,\quad k_{c} =  k_{\text{eq}}\, e^{i \phi_{k}}\,,
\end{split}
\end{align}
so, as the temperature is lowered, the phase is vanishing and the collision appears at pure imaginary frequency and pure real momentum as
\begin{align}\label{}
\begin{split}
\omega_{c} \rightarrow  -i \, \omega_{\text{eq}}\,   \,,\quad k_{c} \rightarrow  k_{\text{eq}}\,.
\end{split}
\end{align}

\paragraph{The upper bound of AdS$_2$ diffusion with magnetic fields.}
{
In holography, it has been studied to find universal bound of the diffusion constant $D$, and it turned out $D$ can have a lower bound~\cite{Blake:2016sud,Blake:2017qgd,Blake:2016jnn,Baggioli:2017ojd,Kim:2017dgz,Ahn:2017kvc}\footnote{This is the case when the IR geometry is AdS$_{2}\times R^2$. There could be a pre-factor when the IR geometry changes.} with the velocity and the time scales related to quantum chaos 
\begin{align}\label{}
\begin{split}
D  \,\ge\,  v_{\text{B}}^2 \, \tau_{\text{L}}  \,,
\end{split}
\end{align}
where $v_{\text{B}}$ is a butterfly velocity and $\tau_{\text{L}}$ is the Lyapunov time. ($v_{\text{B}}, \tau_{\text{L}}$) characterize the chaotic behavior~\cite{Grozdanov:2017ajz} via out-of-time-order correlators (OTOCs)~\cite{larkin1969quasiclassical, Kitaev-2014, Maldacena:2015waa}\footnote{OTOCs is one useful tool to study many-body quantum chaos: $\langle V_0(0) W_{\bf x}(t) V_0(0) W_{\bf x}(t) \rangle \sim 1 + e^{\lambda_{L} \left( t - x/v_{B} \right)}$ where $\lambda_{L}:=1/\tau_{L}$.}.

In recent years, the holographic study of the ``upper" bound also has been investigated with the simple question: the diffusion constant will be also bounded from the above or it just grows forever? It is proposed that the diffusion constant $D$ has the following upper bound~\cite{Arean:2020eus,Wu:2021mkk}:
\begin{align}\label{newbforf6}
\begin{split}
D  \,\le\,  v_{\text{eq}}^2 \, \tau_{\text{eq}}  \,,
\end{split}
\end{align}
where $v_{\text{eq}}:= \omega_{\text{eq}}/k_{\text{eq}}$ and $\tau_{\text{eq}}:=\omega_{\text{eq}}^{-1}$ are the velocity and the timescale associated to the breakdown of hydrodynamics \eqref{EQFOR}. 
The upper bound (an equality) in \eqref{newbforf6} is approached at low $T$ and it can be expressed as 
}
%We can also study the upper bound on the diffusion constant $D$ with the ratio
%
\begin{align}\label{RATIOGOR}
\begin{split}
\frac{D}{v_{\text{eq}}^2 \,\tau_{\text{eq}}}  \,=\,  \frac{k_{\text{eq}}^2}{\omega_{\text{eq}}}\,D \,=\, 1\,.
\end{split}
\end{align}
%
%where $v_{\text{eq}}:= \omega_{\text{eq}}/k_{\text{eq}}$ and $\tau_{\text{eq}}:=\omega_{\text{eq}}^{-1}$ are the velocity and the timescale associated to equilibration. 
%Note that this ratio \eqref{RATIOGOR} is analogues to the lower bound of the diffusion constant $\frac{D}{v_{\text{B}}^2 \,\tau_{\text{L}}} = 1$ in which $v_{\text{B}}$ is a butterfly velocity and $\tau_{\text{L}}$ is a Lyapunov time. \HS{Ref}

\begin{figure}[]
\centering
     {\includegraphics[width=7.2cm]{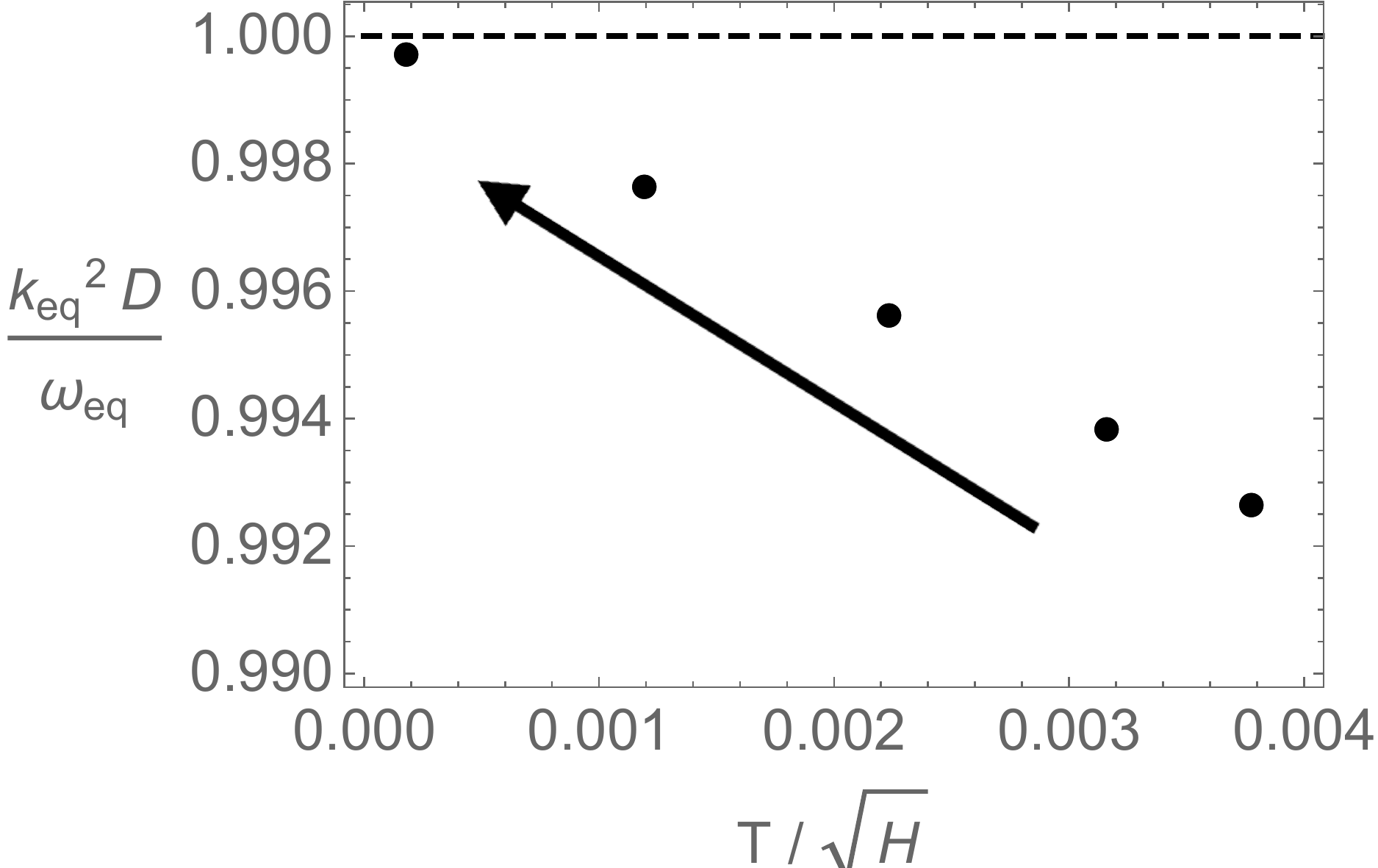} \label{}}
 \caption{The upper bound on the diffusion constant \eqref{DCF}. The dashed line indicates the upper bound \eqref{RATIOGOR}.}\label{Mfig11FIG}
\end{figure}

In Fig. \ref{Mfig11FIG}, we display the conjectured inequality \eqref{newbforf6} still holds in the presence of magnetic fields in which the upper bound \eqref{RATIOGOR} is approached at low $T$.
The appearance of the upper bound implies that we have the following at low $T$  
\begin{align}\label{}
\begin{split}
\omega_{\text{eq}} = D \, k_{\text{eq}}^2 \,,
\end{split}
\end{align}
and it means, as $T$ is lowered, the quasi-normal modes are getting better and better well approximated with i) the hydrodynamic mode \eqref{SD2}; ii) the non-hydrodynamic mode \eqref{irqnms1}.
\begin{figure}[]
\centering
     \subfigure[$H/T^2=10^{5}$]
     {\includegraphics[width=7.2cm]{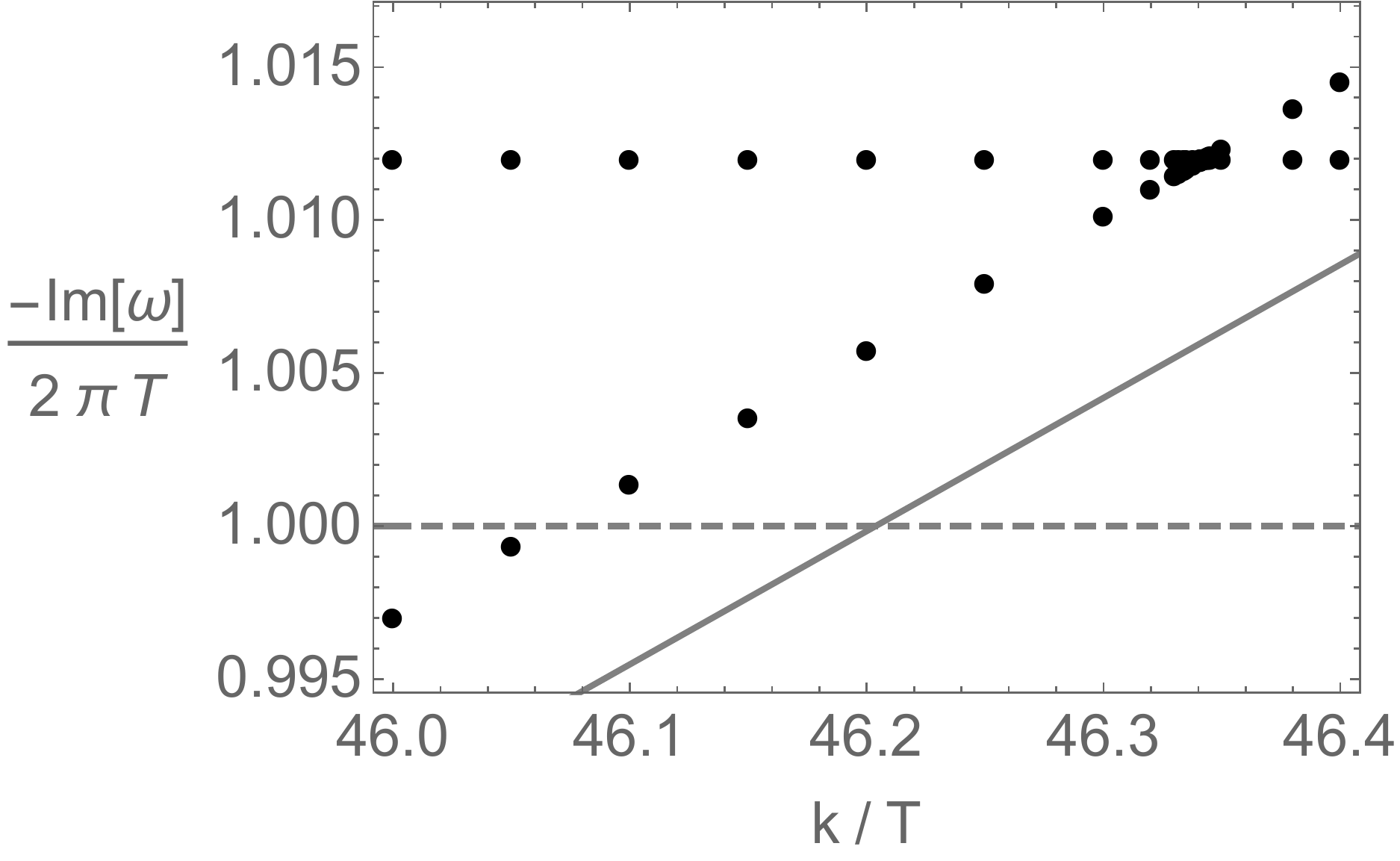} \label{}}
          \subfigure[$H/T^2=3\times10^{7}$]
     {\includegraphics[width=7.2cm]{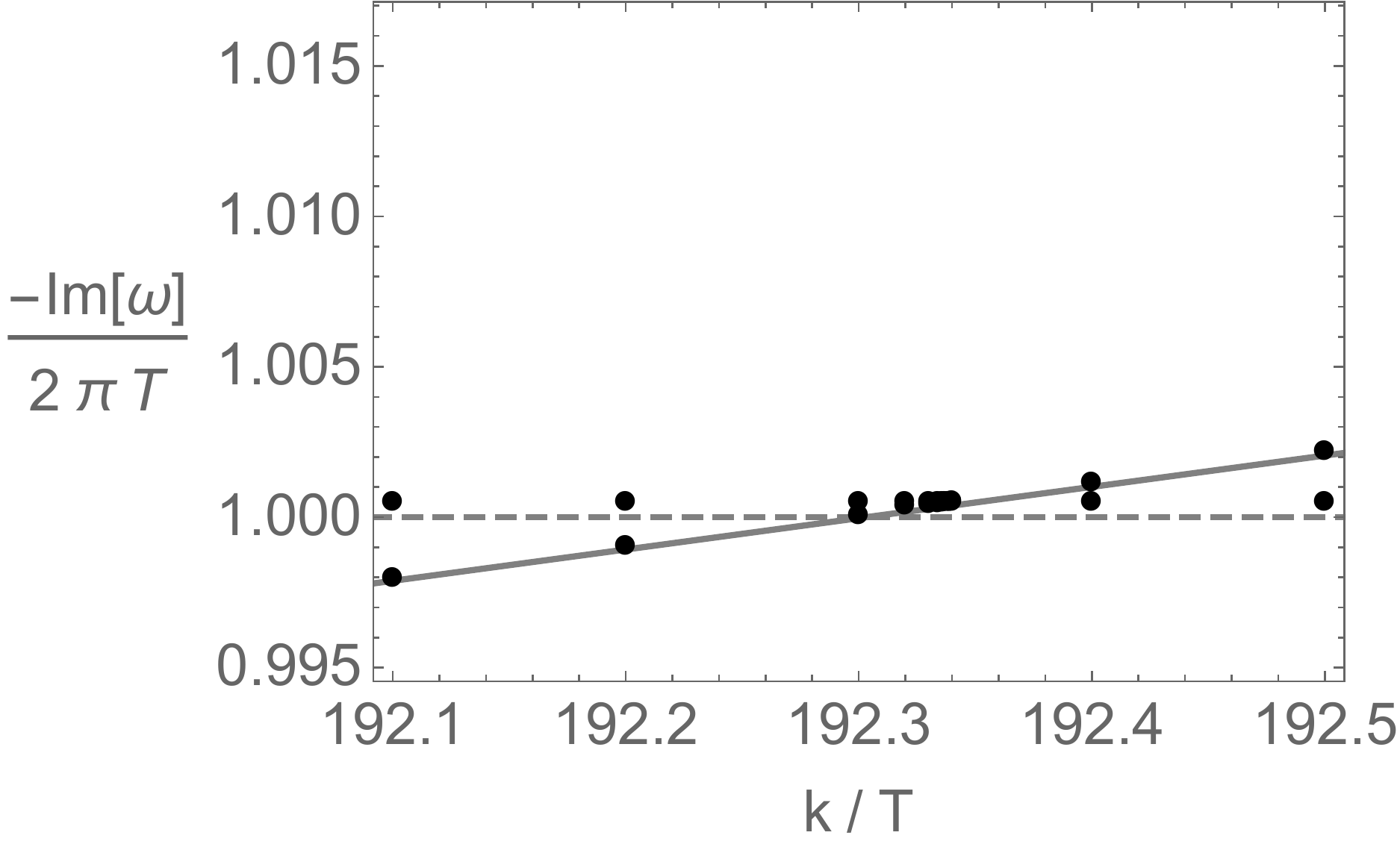} \label{}}
 \caption{Quasi-normal modes (black dots) at $H/T^2=(10^{5}, 3\times10^{7})$ (Left, Right). The solid line is hydrodynamic mode \eqref{SD2} and the dashed line is the first non-hydrodynamic mode $\omega_{0}$ \eqref{irqnms1}.  {Note that the left figure is the gray square region in Fig. \ref{STSKF2}. As $T$ is lowered from (a) to (b), the quasi-normal modes (black dots) are getting better and better well approximated with i) the hydrodynamic mode (solid line); ii) the non-hydrodynamic mode (dashed line).}  }\label{Mfig5FIG}
\end{figure}
See Fig. \ref{Mfig5FIG}.

%%%%%%%%%%%%%%%%%%%%%%%%%%%%%%%%
%    
%%%%%%%%%%%%%%%%%%%%%%%%%%%%%%%%
\section{Comparison with the axion model}

In this section, we compare our results with the previous study of the equilibration data for the energy diffusion~\cite{Arean:2020eus} where the energy diffusion is studied with the linear axion model: \eqref{GENACa} at $N=1$\footnote{The translational invariant system at finite chemical potential~\cite{Arean:2020eus} gives the same result with the linear axion model so we focus on the linear axion model case here.}.
We find two main differences: i) $\Delta(0)$ (or the equilibration time $\omega_{\text{eq}}$); ii) the $T$-scaling in the phase $\phi_{k}$.

\paragraph{The operator of dimension $\Delta(0)$.}
For $\Delta(0)$, the linear axion model gives $\Delta(0)=2$ unlike the finite magnetic field case $\Delta(0)=1$ \eqref{LATdd}.
We may understand this from the perturbation equation in the extremal geometry as
\begin{align} \label{COMPA}
\qquad &\partial_{\zeta}^2 Z_{A} + \left( \frac{2\zeta}{\zeta^2 - \zeta_{h}^2} \right) \partial_{\zeta} Z_{A} + \left(   \frac{c_{1} \, \zeta_{\omega}^2}{(\zeta^2 - \zeta_{h}^2)^2} - \frac{c_{2} \,+ c_{3} \,k^2}{(\zeta^2 - \zeta_{h}^2)} \right) Z_{A} = 0 \,,
\end{align}
where $c_{i}$ are constants depending on the theories:
\begin{align} \label{ddewweqd}
\begin{split}
\text{Magnetic field model:} &\quad c_{1} = \frac{1}{36} \,,\quad c_{2} = 0 \,,\quad c_{3} = \frac{1}{\sqrt{3}H} \,, \\
\text{Linear axion model:} &\quad c_{1} = \frac{1}{9} \,,\,\,\,\quad c_{2} = 2 \,,\quad c_{3} = \frac{2}{m^2} \,,
\end{split}
\end{align}
where the magnetic field model case is given in \eqref{prerq1} and the linear axion model case is in \eqref{prerq1a} with $N=1$.
%\footnote{In appendix \ref{appendixb}, following the similar method in the section 3.1, we explain how to obtain both the perturbation equation in the extremal geometry \eqref{prerq1a} and the IR green's function $\mathcal{G}_{IR}$ \eqref{IRGa} for the axion models for general $N$ \eqref{GENACa}, i.e., this appendix \ref{appendixb} corresponds to the generalization of  \cite{Arean:2020eus} ($N=1$ case) for the general $N$ case. We found $\Delta (0)=2$ for all $N$ \eqref{DDDDDa}, which is consistent with numerical results in \cite{Wu:2021mkk}.}.

Near the AdS$_{2}$ boundary, the equation \eqref{COMPA} with \eqref{bdasd3} gives an operator of dimension $\Delta(k)$ as
\begin{align}\label{}
\begin{split}
\Delta(k) = \frac{1}{2} \left( 1 + \sqrt{(1 + 4 \, c_{2}) + 4 c_{3} \, k^2} \right) \,,
\end{split}
\end{align}
as also can be seen in \eqref{LATdd}, \eqref{LATdda}.
Therefore we have 
\begin{align}\label{deltavalthe}
\begin{split}
\Delta(0) = \frac{1}{2} \left( 1 + \sqrt{(1 + 4 \, c_{2})} \right) \,,
\end{split}
\end{align}
with \eqref{ddewweqd}, which gives rise to $\Delta(0)=1$ for the magnetic field model and $\Delta(0)=2$ for the linear axion model.

\begin{figure}[]
\centering
     \subfigure[$\omega_{\text{eq}}$ vs $T$]
     {\includegraphics[width=5.0cm]{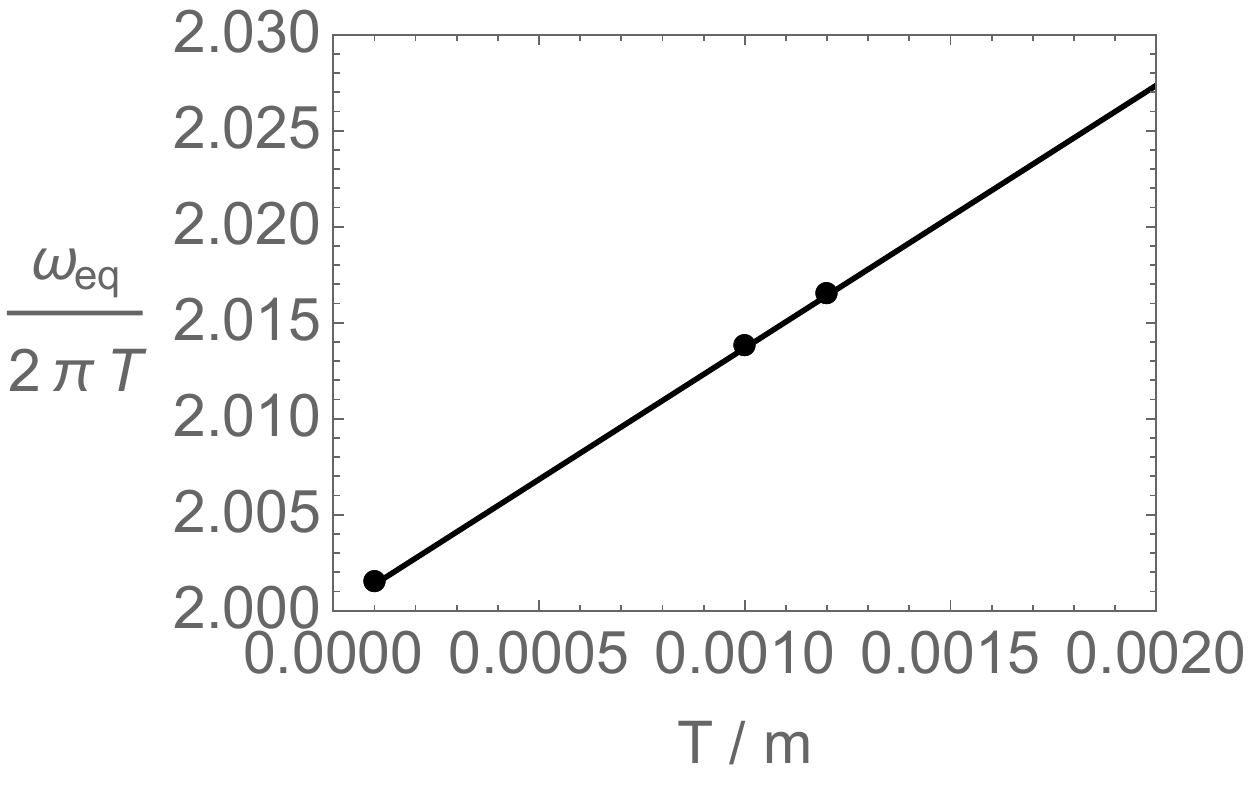} \label{}}
     \subfigure[$k_{\text{eq}}$ vs $T$]
     {\includegraphics[width=4.5cm]{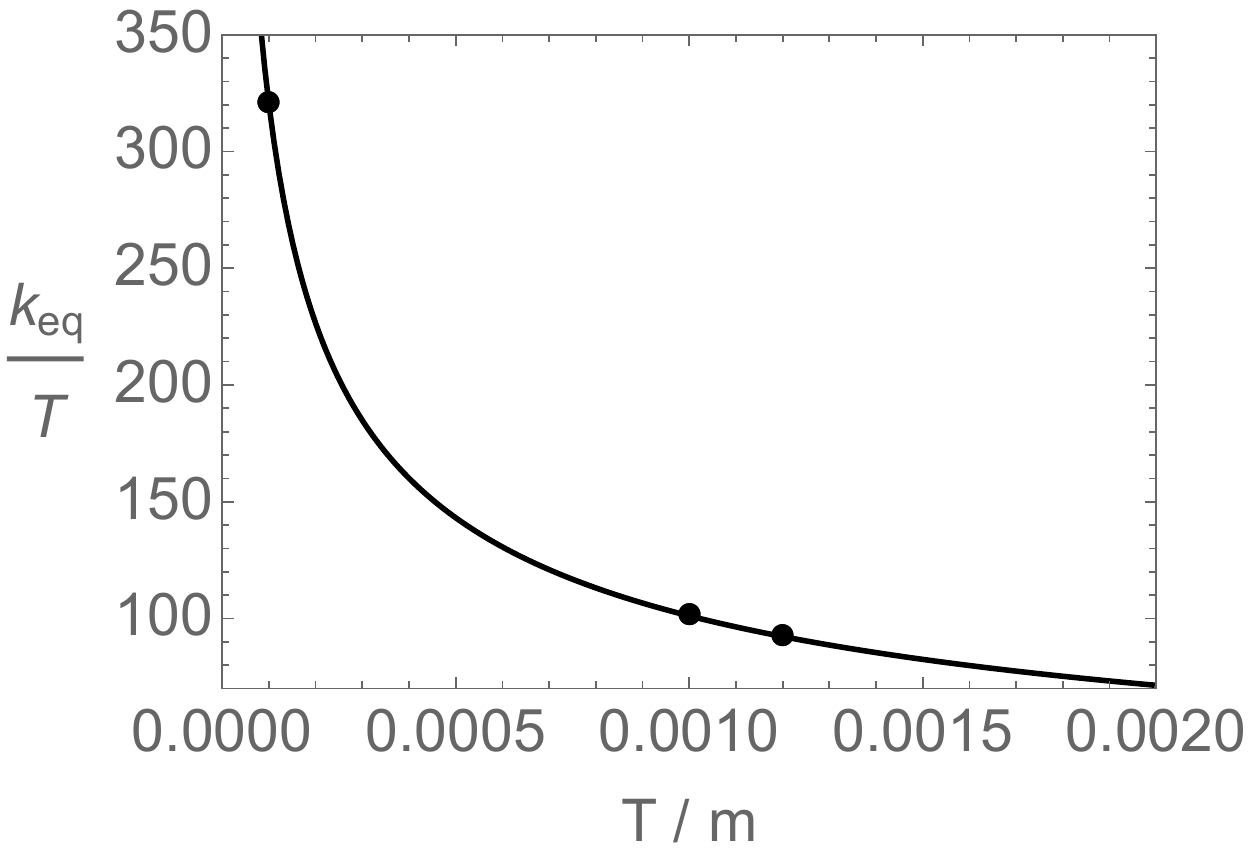} \label{}}
     \subfigure[$\phi_{k}$ vs $T$]
     {\includegraphics[width=4.9cm]{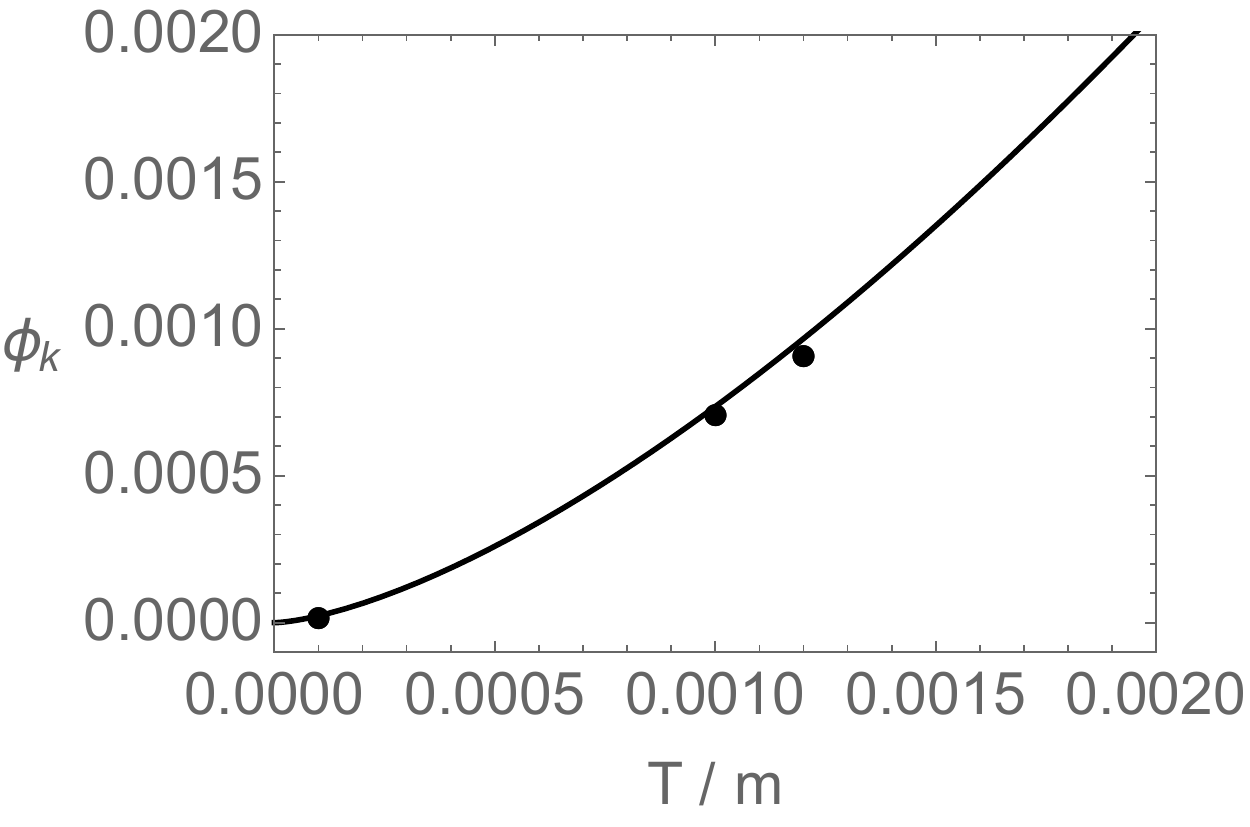} \label{}}
 \caption{The temperature dependence of ($\omega_{\text{eq}},\, k_{\text{eq}}$,\, $\phi_{k}$) for the linear axion model. Dots are numerical data and black solid lines are fitting curves \eqref{FC1a}.}\label{Mfig1111FIG2}
\end{figure}

\paragraph{The low $T$ scaling of the phase $\phi_{k}$:}
Next let us compare \eqref{FC1} with the linear axion model result.
As can be seen in Fig. \ref{Mfig1111FIG2},
the temperature dependence of ($\omega_{\text{eq}},\, k_{\text{eq}}$,\, $\phi_{k}$) for the linear axion model follows 
\begin{align}\label{FC1a}
\begin{split}
\frac{\omega_{\text{eq}}}{2 \pi T} \,\sim\,  \Delta(0) + \# \, T  \,, \qquad \frac{k_{\text{eq}}}{T} \,\sim\,  \frac{\#}{\sqrt{T}} \,, \qquad \phi_{k} \,\sim\,  \# \, T^{3/2}  \,,
\end{split}
\end{align}
with $\Delta(0)=2$, which is consistent with \cite{Arean:2020eus}\footnote{See also analytic treatment for the $T$-scaling \eqref{FC1a} in \cite{Arean:2020eus}.}.
Thus the phase behaves differently in $T$ in the presence of magnetic fields \eqref{FC1}.

Combining \eqref{FC1} with \eqref{FC1a}, we may write the equilibration data of the energy diffusion at low $T$, depending on theories as follows
\begin{align}\label{FC2}
\begin{split}
\omega_{\text{eq}} \,\sim\, 2\pi T \, \Delta(0)  \,, \qquad k_{\text{eq}} \,\sim\,  \sqrt{T} \,, \qquad \phi_{k} \,\sim\, T^{\,7/2 - \Delta(0)}  \,,
\end{split}
\end{align}
where $\Delta(0)$ depends on the theories \eqref{deltavalthe}.

{

\paragraph{Further comments on the operator scaling dimension $\Delta(0)$.}
The axion model \eqref{GENACa}\footnote{For the recent development of this model, see \cite{Baggioli:2021xuv} and references therein.} can be used to study the breaking of translational invariance both explicitly ($N<5/2$) and spontaneously ($N>5/2$) in holography where the diffusion constant $D$ becomes the energy diffusion constant~\cite{Blake:2016sud,Blake:2017qgd,Blake:2016jnn,Kim:2017dgz,Ahn:2017kvc} at $N<5/2$ and the crystal diffusion constant~\cite{Baggioli:2020ljz,Baggioli:2017ojd} at $N>5/2$.

Unlike the explicit symmetry breaking case in \cite{Arean:2020eus} ($N=1$), the operator scaling dimension $\Delta(0)$ has been investigated numerically for the spontaneous symmetry breaking case in \cite{Wu:2021mkk} ($N>5/2$). 
In appendix \ref{appendixb}, following a similar method presented in section 3.1, we showed how to obtain an analytic expression of $\Delta(0)$ of the axion models for general $N$, i.e., the appendix \ref{appendixb} corresponds to the generalization of \cite{Arean:2020eus} ($N=1$ case) to the case of general $N$, and we show that our analytic result is consistent with the numerical results of the previous study~\cite{Wu:2021mkk}.

Let us collect main equations in appendix \ref{appendixb} as follows.
First, the perturbation equation \eqref{prerq1a}  in the extremal geometry for general $N$ is 
\begin{align} \label{prerq1amain}
\qquad &\partial_{\zeta}^2 Z_{A} + \left( \frac{2\zeta}{\zeta^2 - \zeta_{h}^2} \right) \partial_{\zeta} Z_{A} + \left(  \frac{\zeta_{\omega}^2}{9 {N}^2 (\zeta^2 - \zeta_{h}^2)^2} - \frac{ 2\left( 1+ 6^{{\frac{1-N}{N}}} \frac{k^2}{m^2} \right)}{\zeta^2 - \zeta_{h}^2} \right) Z_{A} = 0 \,.
\end{align}
Note that, when $N=1$ this becomes the same equation given in \cite{Arean:2020eus}, which is constructed with the master field.

Then, near the AdS$_{2}$ boundary, the equation \eqref{prerq1amain} with \eqref{bdasd3} gives an operator of dimension $\Delta(k)$ \eqref{LATdda} of the axion models for general $N$ as
\begin{align}\label{}
\begin{split}
\Delta(k) = \frac{1}{2} \left( 1 + \sqrt{9 + 2^{{\frac{2N+1}{N}}}\,3^{{\frac{1-N}{N}}} \frac{k^2}{m^2}} \right)\,,
\end{split}
\end{align}
giving $\Delta (0)=2$ \eqref{DDDDDa} for all $N$, which is consistent with numerical results in \cite{Wu:2021mkk}.
}

%%%%%%%%%%%%%%%%%%%%%%%%%%%%%%%%
%    
%%%%%%%%%%%%%%%%%%%%%%%%%%%%%%%%
\section{Conclusion}
We have studied the breakdown of hydrodynamics in the presence of magnetic fields.
At low $T$, we checked that the equilibration frequency $\omega_{\text{eq}}$ is still associated to the AdS$_{2}$ fixed point and the hydrodynamic mode at the quadratic order \eqref{SD2} is well approximated even beyond $k_{\text{eq}}$, i.e., the equilibration data of the energy diffusion at low $T$ follows 
\begin{align}\label{cc1}
\begin{split}
\omega_{\text{eq}} \,\rightarrow\, 2\pi T \, \Delta(0)  \,, \qquad k_{\text{eq}}^2 \,\rightarrow\,  \frac{\omega_{\text{eq}}}{D}  \,,
\end{split}
\end{align}
where it is first studied with the linear axion model~\cite{Arean:2020eus}.
{
Moreover, we also checked that \eqref{cc1} corresponds to the conjectured upper bound~\cite{Arean:2020eus,Wu:2021mkk} of the diffusion constant $D$, i.e.,
\begin{align}\label{ccccc122}
\begin{split}
D  \,\le\,  \frac{\omega_{\text{eq}}}{k_{\text{eq}}^2}
\,:=\,v_{\text{eq}}^2 \, \tau_{\text{eq}}  \,,
\end{split}
\end{align}
where $v_{\text{eq}}:= \omega_{\text{eq}}/k_{\text{eq}}$ and $\tau_{\text{eq}}:=\omega_{\text{eq}}^{-1}$ are the velocity and the timescale associated to the breakdown of hydrodynamics.
The upper bound (an equality) in \eqref{ccccc122} is approached at low $T$, implying the breakdown of hydrodynamics can be used to find the upper bound of the diffusion constant $D$.

In addition to the magnetic field case we have analytically shown $\Delta(0)=2$ for the axion model independent of the translational symmetry breaking pattern (explicit or spontaneous), which is complementary to previous numerical results~\cite{Arean:2020eus,Wu:2021mkk}.
}

Our work confirmed one of future works proposed in~\cite{Arean:2020eus}: the results \eqref{cc1}-\eqref{ccccc122} continue to hold for AdS$_2$ fixed points supported by a different hierarchy of scales for the energy diffusion (magnetic fields).

\begin{figure}[]
\centering
     {\includegraphics[width=15.2cm]{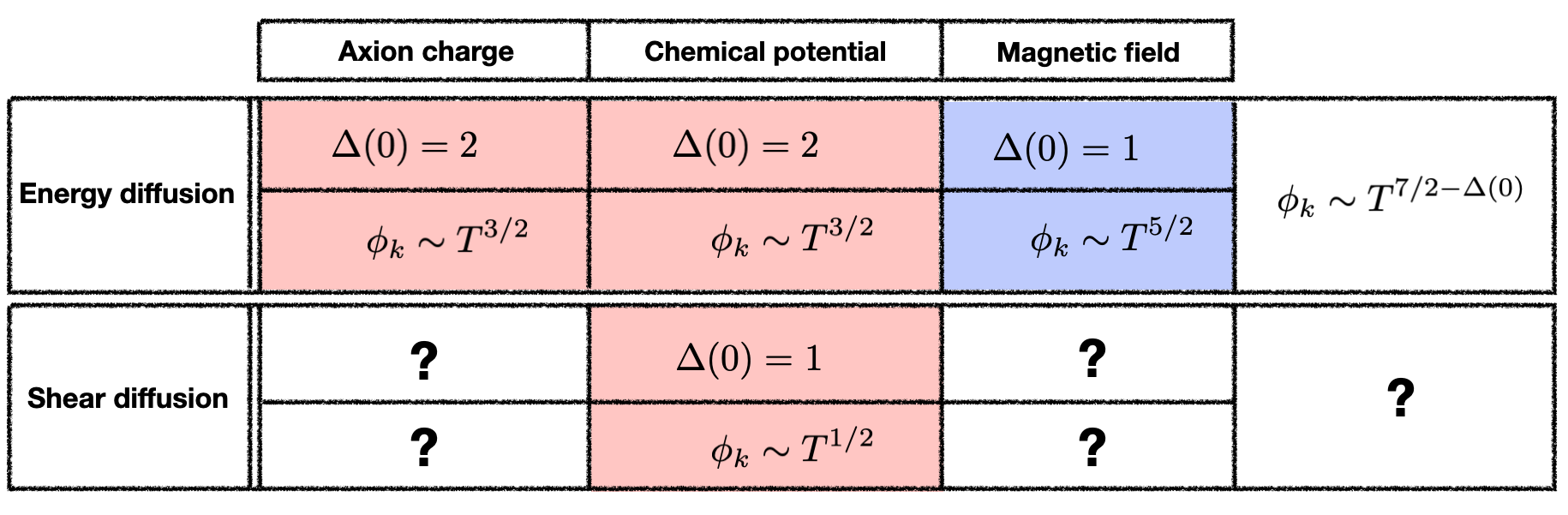} \label{}}
 \caption{The summary for the study of the breakdown of hydrodynamics with ($\omega_{\text{eq}}, k_{\text{eq}}$). Red region is investigated in \cite{Arean:2020eus} and the blue region is considered in this paper.}\label{TBFd1}
\end{figure}

In addition to the theory (or the hierarchy of scales)-independent property \eqref{cc1}-\eqref{ccccc122}, we also found theory-dependent properties in ($\Delta(0)$, $\phi_{k}$) as
\begin{align} \label{}
\begin{split}
\text{Magnetic field model:} &\quad \Delta(0)=1 \,,\quad  \phi_{k} \,\sim\, T^{\,5/2} \,, \\ 
\text{Linear axion model:} &\quad \Delta(0)=2 \,,\quad  \phi_{k} \,\sim\, T^{\,3/2}\,.
\end{split}
\end{align}
Compared with the linear axion model results, we may summarize the low $T$-scaling behavior of $\phi_{k}$ for the energy diffusive hydrodynamic mode as
\begin{align}\label{}
\begin{split}
\phi_{k} \,\sim\, T^{\,7/2 - \Delta(0)}  \,.
\end{split}
\end{align}
It would be interesting to check if this scaling property holds for other value of $\Delta(0)$ with different theories for the energy diffusive hydrodynamics.
Moreover, it will be also interesting to investigate the scaling property of the shear diffusive hydrodynamics{: only the case at finite chemical potential has been investigated so far} \cite{Arean:2020eus}. In other words, we may fill in the empty blanks in Fig. \ref{TBFd1}. In particular, the shear diffusion mode with magnetic fields\footnote{The shear diffusion mode with axion charge may require further analysis due to the absence of the hydrodynamic mode.} may produce interesting features because the hydrodynamic mode is the subdiffusive mode ($\omega = -i D_{\text{shear}} \,k^4$) \eqref{MAGNETO2}.

Another interesting future direction will be to investigate the connection between the pole-skipping and the breakdown of hydrodynamics. 
The pole-skipping point ($\omega_{*}, k_{*}$) \cite{Blake:2018leo,Jeong:2021zhz} is 
\begin{equation}\label{LOWBD}
\begin{split}
\omega_{*} \,=\,  i\, \lambda_{\text{L}} \,, \quad k_{*}^2 \,=\,  -\, \frac{\lambda_{\text{L}}^2}{v_{\text{B}}^2} \,.
\end{split}
\end{equation}
where $\lambda_{\text{L}} = 1/\tau_{L} = 2 \pi T$ is the Lyapunov exponent, $v_{\text{B}}$ butterfly velocity
and this pole-skipping point passes through the energy diffusion mode at low $T$
\begin{equation}\label{qwerqwedd}
\begin{split}
\omega_{*} = -i \,D \,k_{*}^2 \quad\rightarrow\quad D \,=\, i \frac{\omega_{*}}{k_{*}^2} \,=\, \frac{v_{\text{B}}^2}{\lambda_{\text{L}}}  \quad\text{as}\quad T\rightarrow0 \,, 
\end{split}
\end{equation}
where \eqref{LOWBD} is used in the last equality.

At low $T$, we may rewrite the collision point ($\omega_{{c}}, k_{{c}}$) responsible for the breakdown of hydrodynamics as
\begin{equation}\label{UPBD}
\begin{split}
\omega_{c} \,=\,  -i\, \lambda_{\text{L}} \, \Delta(0) \,, \quad k_{c}^2  \,=\, i\, \frac{\omega_{\text{c}}}{D} \,=\,   \frac{\lambda_{\text{L}}^2}{v_{\text{B}}^2} \Delta(0) \,,
\end{split}
\end{equation}
where we can use $D$ \eqref{qwerqwedd} because the temperature is low.
Then, comparing \eqref{LOWBD} with \eqref{UPBD}, it is parametrically possible to make $(\omega_{*}, k_{*}) = (\omega_{c}, k_{c})$ at $\Delta(0)=-1$, which may imply the pole-skipping phenomena might be related to the breakdown of hydrodynamics.

Last but not least, it will be also interesting to figure out some phenomenological effects of the equilibration velocity ($v_{\text{eq}}:= |\omega_{\text{c}}|/|k_{\text{c}}|$), of which the ratio between $v_{\text{eq}}$ and $v_{\text{B}}$ follows
\begin{equation}\label{}
\begin{split}
\frac{v_{eq}}{v_{B}} \,=\, \sqrt{\Delta(0)}\,,
\end{split}
\end{equation}
where \eqref{UPBD} is used.
We leave these subjects as future works and hope to address them in the near future.

%

%%%%%%%%%%%%%%%%%%%%%%%%%%%%%%%%
%    Section: Acknowledgments
%%%%%%%%%%%%%%%%%%%%%%%%%%%%%%%%
\acknowledgments

We would like to thank  Matteo Baggioli, Wei-Jia Li  for valuable discussions and correspondence.  
This work was supported by the National Key R$\&$D Program of China (Grant No. 2018FYA0305800), Project 12035016 supported by National Natural Science Foundation of China, the Strategic Priority Research Program of Chinese Academy of Sciences, Grant No. XDB28000000, Basic Science Research Program through the National Research Foundation of Korea (NRF) funded by the Ministry of Science, ICT $\&$ Future Planning (NRF- 2021R1A2C1006791) and GIST Research Institute(GRI) grant funded by the GIST in 2021.

\appendix
%%%%%%%%%%%%%%%%%%%%%%%%%%%%%%%%
%    
%%%%%%%%%%%%%%%%%%%%%%%%%%%%%%%%

\section{Magneto-hydrodynamics}\label{appendixa}
The magneto hydrodynamics in (2+1) dimensions is given in \cite{Buchbinder:2008dc,Buchbinder:2009aa, Jeong:2021zhz}. For the convenience of the reader, we also present it as follows.

The relevant field theory equation of motions are the conservation  laws:
\begin{align}\label{CONSERVATIONEQ}
\begin{split}
\partial^{\nu} T_{\mu\nu} = F_{\mu\nu} J^{\nu} \,, \qquad
\partial_{\mu} J^{\mu} = 0,
\end{split}
\end{align} 
where $T_{\mu\nu}$ is the stress energy tenser, $J^{\mu}$ is the current and $F_{\mu\nu}$ is the external electromagnetic field.
For the case under consideration, $F_{\mu\nu}$ is taken to be magnetic as
\begin{align}\label{}
\begin{split}
F_{tx} = 0\,, \quad F_{ty} = 0\,, \quad F_{ij} = \epsilon_{ij} H\,,
\end{split}
\end{align} 
where $i,j = (x, y)$.
To first order in derivatives, $T_{\mu\nu}$ and $J^{\mu}$ can be given by the standard expression:
\begin{align}\label{TMUNUEQ}
\begin{split}
T^{\mu\nu} = \epsilon u^{\mu} u^{\nu} + P \Delta^{\mu\nu} - \eta\left(\Delta^{\mu\alpha} \Delta^{\nu\beta}(\partial_{\alpha} u_{\beta} + \partial_{\beta} u_{\alpha})  - \Delta^{\mu\nu} \partial_{\gamma}u^{\gamma} \right) \,,
\end{split}
\end{align} 
where $\Delta^{\mu\nu} = \eta^{\mu\nu} + u^{\mu} u^{\nu}$, $u^{\mu}$ is the fluid 3-velocity, $\epsilon$ and $P$ are the energy density and the pressure respectively, and $\eta$ is the shear viscosity.
%\footnote{As long as we consider the conformal theory, we can set the bulk viscosity $\zeta=0$ in \eqref{TMUNUEQ}.}.
Similarly, the current is given by
\begin{align}\label{JCURRENTEQ}
\begin{split}
J^{\mu} = \rho u^{\mu} + \sigma_{Q} \Delta^{\mu\nu} (-\partial_{\nu} \mu + F_{\nu\alpha} u^{\alpha} + \frac{\mu}{T}\partial_{\nu} T) \,,
\end{split}
\end{align} 
where $\rho$ is the charge density, $\mu$ is the chemical potential, $T$ is the temperature and $\sigma_{Q}$ is the conductivity coefficient.
For the study of fluctuation around the equilibrium in which
\begin{align}\label{}
\begin{split}
u^{\mu} = (1, 0, 0) \,, \qquad T = \text{constant}\,, \qquad \mu = \text{constant} \,,
\end{split}
\end{align} 
we choose $(\delta u_{x}, \, \delta u_{y},\, \delta T,\, \delta \mu)$ as the independent variables. Then, with the plane wave form $e^{-i \omega t + i k x}$, one can find the eight relevant fluctuations: 
\begin{align}\label{FLUCFLUC}
\begin{split}
\delta T^{tt}\,,\,\,  \delta T^{tx}\,, \,\, \delta T^{ty}\,,\,\,  \delta T^{xy}\,, \,\, \delta T^{xx}\,, \,\, 
\delta J^{t}\,, \,\, \delta J^{x}\,, \,\, \delta J^{y}\,.
\end{split}
\end{align} 
For the specific form of \eqref{FLUCFLUC}, see equation (2.14) and (2.15) in \cite{Buchbinder:2008dc}. 
After putting all the fluctuations \eqref{FLUCFLUC} into the equation \eqref{CONSERVATIONEQ} and performing a Fourier transformation, we can get the four coupled equations:
\begin{align}\label{}
\begin{split}
 0 &= \omega \left(  \left(\frac{\partial{\epsilon}}{\partial{\mu}}\right)_{T}  \delta \mu +   \left(\frac{\partial{\epsilon}}{\partial{T}}\right)_{\mu} \delta T    \right) - k (\epsilon + P) \delta u_{x} \,, \\
 0 &= \omega (\epsilon + P) \delta u_{x} - k \left(  \left(\frac{\partial{P}}{\partial{\mu}}\right)_{T}  \delta \mu +   \left(\frac{\partial{P}}{\partial{T}}\right)_{\mu} \delta T    \right) + i k^2 \eta \, \delta u_{x} + i \sigma_{Q} H^2 \delta u_{x} + i H \rho \delta u_{y} \,, \\
 0 &= \omega (\epsilon + P) \delta u_{y} - k H \sigma_{Q} \left(\delta \mu - \frac{\mu}{T} \delta T\right) - i H \rho \delta u_{x} + i \sigma_{Q} H^2 \delta u_{y} + i k^2 \eta \delta u_{y} \,, \\
 0 &= \omega \left(  \left(\frac{\partial{\rho}}{\partial{\mu}}\right)_{T}  \delta \mu +   \left(\frac{\partial{\rho}}{\partial{T}}\right)_{\mu} \delta T    \right) - k \rho \delta u_{x} + k \sigma_{Q} H \delta u_{y} + i k^2 \sigma_{Q} \left(\delta\mu - \frac{\mu}{T}\delta T \right) \,.
\end{split}
\end{align} 
Although these equations are all coupled with each other, if we consider  zero charge density($\rho=0$) and no chemical potential($\mu=0$) which is the same condition\footnote{In addition to zero charge condition, motivated by M2-brane magneto hydrodynamics, we may set $\left(\frac{\partial \rho}{\partial T}\right)_{\mu} = \left(\frac{\partial \epsilon}{\partial \mu}\right)_{T} = 0$, $\left(\frac{\partial \rho}{\partial \mu}\right)_{T} \neq 0$, $\left(\frac{\partial \epsilon}{\partial T}\right)_{\mu} \neq 0$.} used for the holographic computation in this paper, these equations can be simplified and decoupled into two independent pairs: i) sound channel; ii) shear channel.

\paragraph{Sound channel:}
The first pair of equation is called sound channel and it reads as
\begin{align}\label{SOUNDCHNNEL}
\begin{split}
&\omega \left(\frac{\partial\epsilon}{\partial T}\right)_{\mu} \delta T - k(\epsilon + P) \delta u_{x} = 0\,, \\
& \omega(\epsilon + P)\delta u_{x} - k\left(\frac{\partial P}{\partial T}\right)_{\mu} \delta T + i k^2 \eta \delta u_{x} + i \sigma_{Q} H^2 \delta u_{x} = 0 \,.
\end{split}
\end{align} 
Combining these two equations, we can obtain one equation about $\omega$, which is a second order equation for $\omega$. Then, by solving it, we get two dispersion relations $\omega(k)$.
When $H=0$, the equation gives the sound mode with the dispersion relation\footnote{We correct the typo in \cite{Buchbinder:2008dc,Buchbinder:2009aa}: there should be factor $1/2$ in $k^2$ order.}:
\begin{align}\label{SCDSP1}
\begin{split}
(H=0): \quad \omega = \pm \sqrt{\frac{\partial P}{\partial \epsilon}} k \,-\, i  \frac{\eta}{2(\epsilon+P)} \,k^2  \,+\, \mathcal{O}(k^3)\,.
\end{split}
\end{align} 
On the other hand, when the magnetic field is turned on, the equation gives the dramatically changed dispersion relations:
\begin{align}\label{SCDSP2}
\begin{split}
(H\neq0): \quad \omega = -i \frac{\sigma_{Q} H^2}{\epsilon+P} + \mathcal{O}(k^2) \,, \qquad
\omega = -i  \frac{\partial P}{\partial \epsilon} \frac{\epsilon + P}{\sigma_{Q} H^2} \,k^2 + \mathcal{O}(k^4) \,.
\end{split}
\end{align} 
As it has been explained in \cite{Buchbinder:2008dc,Buchbinder:2009aa}, this drastic change occurs because, as one can see from \eqref{SOUNDCHNNEL}, the small $H$ limit does not commute with the hydrodynamic limit of small $\omega$ and $k$. Thus, the magnetic field cannot be considered as a small perturbation.

\paragraph{Shear channel:}
The second pair of equations is called the shear channel and it gives
\begin{align}\label{SHEARCHNNEL}
\begin{split}
&\omega(\epsilon+P)\delta u_{y} - k H \sigma_{Q} \delta {\mu} + i \sigma_{Q} H^2 \delta u_{y} + i k^2 \eta \delta u_{y} = 0 \,, \\
&\omega \left( \frac{\partial \rho}{\partial \mu} \right)_{T} \delta \mu + k \sigma_{Q} H \delta u_{y} + i k^2 \sigma_{Q} \delta \mu = 0 \,.
\end{split}
\end{align} 
When $H=0$, these equations give two diffusive modes:
\begin{align}\label{MAGNETO1}
\begin{split}
(H=0): \quad \omega = -i  \frac{\sigma_{Q}}{\left(\frac{\partial\rho}{\partial\mu}\right)_{T}} \,k^2  \,, \qquad
\omega = -i \frac{\eta}{\epsilon+P} k^2 \,,
\end{split}
\end{align} 
where the first (second) one is called the charge (shear) diffusion mode. 
In the presence of magnetic field, these diffusive modes also undergo a dramatic change:
\begin{align}\label{MAGNETO2}
\begin{split}
(H\neq0): \quad \omega = -i \frac{\sigma_{Q} H^2}{\epsilon+P} + \mathcal{O}(k^2) \,, \qquad
\omega = -i  \frac{\eta}{H^2 \left( \frac{\partial \rho}{\partial \mu} \right)_{T}} \,k^4 + \mathcal{O}(k^6) \,.
\end{split}
\end{align} 
%

%%%%%%%%%%%%%%%%%%%%%%%%%%%%%%%%%%%%%%
%    
%%%%%%%%%%%%%%%%%%%%%%%%%%%%%%%%%%%%%%
\section{Axion model}\label{appendixb}

\subsection{Holographic setup}

We consider the axion model in (3+1) dimensions as 
\begin{equation}
\begin{split}
S = \int \dd^4x \sqrt{-g} \,\left( R \,+\, 6 \,-\, X^{N} \right) \, \label{GENACa}
\end{split}
\end{equation}
where
\begin{align}\label{}
\begin{split}
X := \frac{1}{2}\sum_{i=1}^{2}\left(\partial\varphi_i\right)^2 \,, \qquad \varphi_i = m \, x^i.
\end{split}
\end{align}

The background metric is given as
\begin{equation}\label{bgmora}
\begin{split}
\dd s^2 =  -f(r)\, \dd t^2 +  \frac{1}{f(r)} \, \dd r^2  + r^2 (\dd x^2 + \dd y^2) \,,
\end{split}
\end{equation}
where
\begin{equation}\label{BGMa}
\begin{split}
 f(r)\,= r^2 - \frac{m_{0}}{r} \,+ \, \frac{m^{2N}}{2\,(2N-3)\,r^{2N-2}} \,, \quad m_{0} = r_{h}^3\left( 1 +  \frac{m^{2N}}{(4N-6)\, r_{h}^{2N}} \right) \,,
\end{split}
\end{equation}
here $m_{0}$ is determined by the condition $f(r_{h})=0$.

The temperature reads
\begin{align}\label{THMERRESULTSGENa}
\begin{split}
 T &\,=\, \frac{1}{4\pi} \left( 3\,r_{h} \,-\, \frac{m^{2N}}{2\,r_{h}^{2N-1}}  \right) \,.
\end{split}
\end{align}

Based on the background geometry, we study the longitudinal sector as
\begin{align}\label{}
\begin{split}
\delta g_{tt} &= h_{tt}(r) \,e^{-i \, \omega \, t + i \, k \, x} \,,\, \quad  \delta g_{tx} = h_{tx}(r) \,e^{-i \, \omega \, t + i \, k \, x} \,,  \quad
\delta g_{xx} = h_{xx}(r) \,e^{-i \, \omega \, t + i \, k \, x} \,, \\
\delta g_{yy} &= h_{yy}(r) \,e^{-i \, \omega \, t + i \, k \, x} \,, \quad\delta \varphi_{x} = \psi_{x}(r) \,e^{-i \, \omega \, t + i \, k \, x} \,.
\end{split}
\end{align} 
%

%%%%%%%%%%%%%%%%%%%%%%%%%%%%%%%%%%%%%%
%    
%%%%%%%%%%%%%%%%%%%%%%%%%%%%%%%%%%%%%%
\subsection{Gauge-invariant perturbations}

We choose the following deffeomorphism and gauge-invariant combinations:
\begin{align}
\begin{split}
Z_{H} &:= \frac{4 k}{\omega} \,  h_{t}^{x} \,+\,  2  h_{x}^{x} - \left( 2 - \frac{k^2}{\omega^2}\frac{f'(r)}{r} \right)  h_{y}^{y}  + \frac{2k^2}{\omega^2}\frac{f(r)}{r^2}  h_{t}^{t}  \,, \\
Z_{A} &:= \psi_{x}  \,+\, \frac{i m}{2k} \left( h_{x}^{x}- h_{y}^{y}\right) \,,
\end{split}
\end{align}
where we raised an index on fluctuation fields using the background metric \eqref{BGMa}.
Then, we can obtain the second order equations for $Z_{H}$ and $Z_{A}$ of the following form\footnote{In \eqref{ZAZHEOMa}, there are two equations. Substituting the second equation into the first equation, one can obtain a single equation of motion for $Z_{A}$ when $N=1$.}:
\begin{align}\label{ZAZHEOMa}
\begin{split}
&0 \,=\, A_{H}\,Z_{H}''  \,+\, B_{H}\,Z_{H}' \,+\, C_{H}\,Z_{A}' \,+\, D_{H}\,Z_{H}  \,+\, E_{H}\,Z_{A} \,, \\
&0 \,=\, A_{A}\,Z_{A}''  \,\,+\, B_{A}\,Z_{H}'' \,\,+\, C_{A}\,Z_{H}'\,\,\,+\, D_{A}\,Z_{A}' \,\,+\, E_{A}\,Z_{H}  \,\,+\, F_{A}\,Z_{A} \,.
\end{split}
\end{align}
Since the coefficients of equations are lengthy and cumbersome we will not write them out in this paper.

Next, we solve the equations of motion \eqref{ZAZHEOMa} with two boundary conditions: one from incoming boundary condition at the horizon and the other from the AdS boundary.
First, near the horizon ($r\rightarrow r_{h}$) the variables are expanded as 
\begin{align}\label{APPENHORIZONa}
\begin{split}
Z_{H} = (r-r_{h})^{\nu_{\pm}} \left( Z_{H}^{(I)} \,+\, Z_{H}^{(II)} (r-r_{h}) \,+\, \dots   \right ) \,, \\
Z_{A} = (r-r_{h})^{\nu_{\pm}} \left( Z_{A}^{(I)} \,+\, Z_{A}^{(II)} (r-r_{h}) \,+\, \dots   \right ) \,.
\end{split}
\end{align}
where $\nu_{\pm}:= \pm i\omega/4 \pi T$ and we choose $\nu_{-}:= - i\omega/4 \pi T$ for the incoming boundary condition.
After plugging \eqref{APPENHORIZONa} into equations \eqref{ZAZHEOMa}, one can find that higher order horizon coefficients are determined by two independent horizon variables : $(Z_{H}^{(I)}, Z_{A}^{(I)})$.

Near the AdS boundary ($r\rightarrow \infty$), the asymptotic behavior of solutions depends on the symmetry breaking patterns. 
\paragraph{Example 1: the explicit symmetry breaking case ($N=1$).} For the explicit symmetry breaking, the asymptotic behavior of solutions is
\begin{align}\label{}
\begin{split}
&Z_{H} = Z_{H}^{(S)} \, r^{0} \,(1 \,+\, \dots) \,+\, Z_{H}^{(R)} \, r^{-3} \,(1 \,+\, \dots) \,, \\
&Z_{A} = Z_{A}^{(S)} \, r^{0} \,(1 \,+\, \dots) \,+\, Z_{A}^{(R)}\, r^{-3} \,(1 \,+\, \dots) \,,
\end{split}
\end{align}
where the superscripts mean that $(S)$ is the source term and $(R)$ is a response term.

\paragraph{Example 2: the spontaneous symmetry breaking case ($N=3$).}
For the spontaneous symmetry breaking, they expand as 
\begin{align}\label{}
\begin{split}
&Z_{H} = Z_{H}^{(S)} \, r^{0} \,(1 \,+\, \dots) \,+\, Z_{H}^{(R)} \, r^{-3} \,(1 \,+\, \dots) \,, \\
&Z_{A} = Z_{A}^{(S)} \, r^{1} \,(1 \,+\, \dots) \,+\, Z_{A}^{(R)}\, r^{0} \,(1 \,+\, \dots)  \,.
\end{split}
\end{align}

\paragraph{The determinant method for the quasi-normal modes:}
We can compute the quasi-normal modes by employing the determinant method. Using the shooting method, we can construct the matrix of sources with the source terms near the boundary:
\begin{align}\label{APPENSMATAa}
\begin{split}
S = \left(\begin{array}{cc}Z_{H}^{(S)(I)} & Z_{H}^{(S)(II)} \\Z_{A}^{(S)(I)} & Z_{A}^{(S)(II)}\end{array}\right) \,,
\end{split}
\end{align}
where the $S$-matrix is $2\times2$ matrix because we can get two independent solutions with two independent shooting variables at the horizon \eqref{APPENHORIZONa}. Note that $I (II)$ in \eqref{APPENSMATAa} denotes that the source terms are obtained by the $I (II)$ th shooting.

Finally, the quasi-normal mode spectrum, the dispersion relation in which the holographic Green's functions have a pole, can be given by the value of ($\omega, k$) for which the determinant of $S$-matrix \eqref{APPENSMATAa} vanishes.

%%%%%%%%%%%%%%%%%%%%%%%%%%%%%%%%%%%%%%
%    
%%%%%%%%%%%%%%%%%%%%%%%%%%%%%%%%%%%%%%
\subsection{Near-horizon perturbation equations}

In this section, we study the perturbation equation \eqref{ZAZHEOMa} in the $\text{AdS}_2 \times \text{R}^2$ spacetime that emerges near the horizon at low temperatures.
We will closely follow the method given in the appendix A in \cite{Arean:2020eus} and we will show that our computations reproduce results in \cite{Arean:2020eus} when $N=1$. This section can be considered as the generalization of \cite{Arean:2020eus}.

\paragraph{The extremal geometry of the axion model.}
Let us first discuss the small temperature condition. From \eqref{THMERRESULTSGENa}, one can check the temperature becomes zero at 
\begin{align}\label{}
\begin{split}
r_{h} = r_{e} \,,\quad  r_{e} := \frac{m}{6^{\frac{1}{2N}}}\,.
\end{split}
\end{align}
Thus, by putting the following relation into \eqref{THMERRESULTSGENa}
\begin{align}\label{}
\begin{split}
 T \,=\, 0 \,+\, \epsilon \, \delta T \,, \quad r_{h} \,=\, r_{e} \,+\, \epsilon \, \zeta_{h} \,,
\end{split}
\end{align}
and take $\epsilon\rightarrow0$, we can find the small temperature correction $\zeta_{h}$ as
\begin{align}\label{}
\begin{split}
\zeta_{h} \,=\, \frac{2\pi}{3N} \delta T \,.
\end{split}
\end{align}

Next, in order to obatin the extremal geometry, one can consider the following coordinate transformation \cite{Arean:2020eus,Faulkner:2009wj} in \eqref{bgmora}
\begin{align}\label{COTa}
\begin{split}
 r \,=\, r_{e} \,+\, \epsilon \, \zeta \,, \quad r_{h} \,=\, r_{e} \,+\, \epsilon \, \zeta_{h} \,,\quad t \,=\, \frac{u}{\epsilon} \,,
\end{split}
\end{align}
and take $\epsilon\rightarrow0$. Then we can change the coordinate from ($t, r$) to ($u, \zeta$) as follows
\begin{equation}\label{EXTGa}
\begin{split}
\dd s^2 =  -\frac{\zeta^2}{L^2} \left( 1-\frac{\zeta_{h}}{\zeta} \right)^2 \, \dd u^2 +  \frac{L^2}{\zeta^2 \, \left( 1-\frac{\zeta_{h}}{\zeta} \right)^2} \, \dd \zeta^2  + r_{e}^2 (\dd x^2 + \dd y^2) \,,
\end{split}
\end{equation}
where the AdS$_{2}$ radius of curvature is $L^2=1/(3N)$. Note that this geometry corresponds to the $\text{AdS}_2 \times \text{R}^2$ geometry with the small temperature correction ($\zeta_{h}$), i.e., $\zeta$ runs from $\zeta = \zeta_{h}$ to $\zeta = \infty$ (the $\text{AdS}_{2}$ boundary).

\paragraph{The perturbation equation in the extremal geometry:}

Using the following coordinate transformation,
\begin{align}\label{COT2a}
\begin{split}
 r \,=\, r_{e} \,+\, \epsilon \, \zeta \,, \quad r_{h} \,=\, r_{e} \,+\, \epsilon \, \zeta_{h} \,,\quad \omega \,=\, \epsilon \, \zeta_{\omega}  \,,
\end{split}
\end{align}
we can obtain the perturbation equations \eqref{ZAZHEOMa} in the extremal geometry \eqref{EXTGa} in $\epsilon\rightarrow0$ limit.
Note that \eqref{COT2a} is the Fourier space version of \eqref{COTa}.

At the leading order in $\epsilon$, one can check only the second perturbation equation in \eqref{ZAZHEOMa} survives: the first perturbation equation starts at $\mathcal{O}(\epsilon)$ order.
In particular, the survived equation of motion is composed of only one field $Z_{A}(\zeta)$:
\begin{align} \label{prerq1a}
\qquad &\partial_{\zeta}^2 Z_{A} + \left( \frac{2\zeta}{\zeta^2 - \zeta_{h}^2} \right) \partial_{\zeta} Z_{A} + \left(  \frac{\zeta_{\omega}^2}{9 {N}^2 (\zeta^2 - \zeta_{h}^2)^2} - \frac{ 2\left( 1+ 6^{{\frac{1-N}{N}}} \frac{k^2}{m^2} \right)}{\zeta^2 - \zeta_{h}^2} \right) Z_{A} = 0 \,.
\end{align}
Note that, when $N=1$ this becomes the same equation given in \cite{Arean:2020eus}, which is constructed with the master field.

Near the AdS$_{2}$ boundary ($\zeta \rightarrow \infty$), the solution is expanded as 
\begin{align}\label{bdasd3a}
\begin{split}
Z_{A} \,=\, Z^{(S)} \, \zeta^{\Delta(k)-1}  + Z^{(R)} \, \zeta^{-\Delta(k)}  \,,
\end{split}
\end{align}
where $\Delta(k)$ depends on the value of $N$ as 
\begin{align}\label{LATdda}
\begin{split}
\Delta(k) = \frac{1}{2} \left( 1 + \sqrt{9 + 2^{{\frac{2N+1}{N}}}\,3^{{\frac{1-N}{N}}} \frac{k^2}{m^2}} \right).
\end{split}
\end{align}

Then, the infra-red Green's function can be found explicitly by solving the equations \eqref{prerq1a} and imposing the usual AdS/CFT rules\footnote{The general solution has the associated Legendre function of the second kind which corresponds to the out-going solution at $\zeta = \zeta_{h}$, so we should discard it.} at the AdS$_{2}$ boundary ($\zeta \rightarrow \infty$) \cite{Faulkner:2009wj,Hartnoll:2012rj} 
\begin{align}
\begin{split}
\mathcal{G}_{IR} \propto \frac{Z^{(R)}}{Z^{(S)}} \,,
\end{split}
\end{align}
where  $Z^{(S)}$, $Z^{(R)}$ are coefficients in \eqref{bdasd3a}.
Let us show the explicit form of the infra-red Green's function 
\begin{align}\label{IRGa}
\begin{split}
\mathcal{G}_{IR} =   \frac{2\Delta(k)-1}{2} \left(\frac{3{N}}{\pi}\right)^{1-2\Delta(k)}    T^{2 \, \Delta (k) -1}  \frac{ \Gamma \left(  \frac{1}{2} - \Delta(k) \right) \Gamma \left(  \Delta(k)  - \frac{i\omega}{2\pi  T}\right) }{ \Gamma \left(  \frac{1}{2} + \Delta(k) \right) \Gamma \left( 1- \Delta(k)  - \frac{i\omega}{2\pi  T}\right) } 
\end{split}
\end{align}
where when $N=1$ this becomes the same infra-red Green's function \cite{Arean:2020eus}\footnote{We also replace notations: $\delta T \rightarrow T$, $\zeta_{\omega} \rightarrow \omega$.}. 
Thus, we showed that the axion parameter {$N$} can change two things in $\mathcal{G}_{IR}$: i) the overall pre-factor; ii) $\Delta (k)$ \eqref{LATdda}.

From \eqref{IRGa}, at any zon-zero temperature the infra-red Green's function has a series of poles along the negative imaginary frequency axis at the locations
\begin{align}\label{}
\begin{split}
\omega_{n} \,=\, -i \, 2\pi\, T \left( n \,+\, \Delta (k) \right)\,, \qquad n \,=\, 0,1,2,\dots \,.
\end{split}
\end{align}
In order to reconstruct $G_{\mathcal{\epsilon}\mathcal{\epsilon}}$ (the energy density green's function) analytically, one must extend the near-horizon solutions through the rest of the spacetime, which is difficult to do. However, it is enough to observe that $G_{\mathcal{\epsilon}\mathcal{\epsilon}}$ exhibits poles where its locations approach those of the poles of $\mathcal{G}_{IR}$ as $k, T \rightarrow 0$~\cite{Arean:2020eus}. Specifically, this means that in this limit $G_{\mathcal{\epsilon}\mathcal{\epsilon}}$ exhibits poles at 
\begin{align}\label{}
\begin{split}
\omega_{n} \,=\, -i \, 2\pi\, T \left( n \,+\, \Delta (0) \right)\,, \qquad n \,=\, 0,1,2,\dots \,.
\end{split}
\end{align}
%\

Therefore, we expect the equilibration frequency ($\omega_{\text{eq}} \,:=\,  \omega_{n=0}$) for general $N$ would be 
\begin{align}\label{DDDDDa}
\begin{split}
 \frac{\omega_{\text{eq}}}{2 \pi T} \,=\, - i \, \Delta(0) \,=\, - 2 i \,,
\end{split}
\end{align}
where from \eqref{LATdda}, $\Delta (0)=2$ for all $N$.

%%%%%%%%%%%%%%%%%%%%%%%%%%%%%%%%
%    Section: END
%%%%%%%%%%%%%%%%%%%%%%%%%%%%%%%%

%\bibliography{HyunSikRefs}

\begin{thebibliography}{10}

\bibitem{LLDM87}
L.~D. Landau and E.~M. Lifshitz, \emph{{Fluid Mechanics}}, {\emph{Pergamon
  Press} (1987) }.

\bibitem{Withers:2018srf}
B.~Withers, \emph{{Short-lived modes from hydrodynamic dispersion relations}},
  \href{http://dx.doi.org/10.1007/JHEP06(2018)059}{\emph{JHEP} {\bf 06} (2018)
  059}, [\href{http://arxiv.org/abs/1803.08058}{{\tt 1803.08058}}].

\bibitem{Grozdanov:2019kge}
S.~Grozdanov, P.~K. Kovtun, A.~O. Starinets and P.~Tadi\'c, \emph{{Convergence
  of the Gradient Expansion in Hydrodynamics}},
  \href{http://dx.doi.org/10.1103/PhysRevLett.122.251601}{\emph{Phys. Rev.
  Lett.} {\bf 122} (2019) 251601}, [\href{http://arxiv.org/abs/1904.01018}{{\tt
  1904.01018}}].

\bibitem{Grozdanov:2019uhi}
S.~Grozdanov, P.~K. Kovtun, A.~O. Starinets and P.~Tadi\'c, \emph{{The complex
  life of hydrodynamic modes}},
  \href{http://dx.doi.org/10.1007/JHEP11(2019)097}{\emph{JHEP} {\bf 11} (2019)
  097}, [\href{http://arxiv.org/abs/1904.12862}{{\tt 1904.12862}}].

\bibitem{Heller:2020hnq}
M.~P. Heller, A.~Serantes, M.~Spali\'nski, V.~Svensson and B.~Withers,
  \emph{{Convergence of hydrodynamic modes: insights from kinetic theory and
  holography}},  \href{http://arxiv.org/abs/2012.15393}{{\tt 2012.15393}}.

\bibitem{Heller:2020uuy}
M.~P. Heller, A.~Serantes, M.~Spali\'nski, V.~Svensson and B.~Withers,
  \emph{{The hydrodynamic gradient expansion in linear response theory}},
  \href{http://arxiv.org/abs/2007.05524}{{\tt 2007.05524}}.

\bibitem{Heller:2013fn}
M.~P. Heller, R.~A. Janik and P.~Witaszczyk, \emph{{Hydrodynamic Gradient
  Expansion in Gauge Theory Plasmas}},
  \href{http://dx.doi.org/10.1103/PhysRevLett.110.211602}{\emph{Phys. Rev.
  Lett.} {\bf 110} (2013) 211602}, [\href{http://arxiv.org/abs/1302.0697}{{\tt
  1302.0697}}].

\bibitem{Abbasi:2020ykq}
N.~Abbasi and S.~Tahery, \emph{{Complexified quasinormal modes and the
  pole-skipping in a holographic system at finite chemical potential}},
  \href{http://dx.doi.org/10.1007/JHEP10(2020)076}{\emph{JHEP} {\bf 10} (2020)
  076}, [\href{http://arxiv.org/abs/2007.10024}{{\tt 2007.10024}}].

\bibitem{Jansen:2020hfd}
A.~Jansen and C.~Pantelidou, \emph{{Quasinormal modes in charged fluids at
  complex momentum}},
  \href{http://dx.doi.org/10.1007/JHEP10(2020)121}{\emph{JHEP} {\bf 10} (2020)
  121}, [\href{http://arxiv.org/abs/2007.14418}{{\tt 2007.14418}}].

\bibitem{Grozdanov:2020koi}
S.~Grozdanov, \emph{{Bounds on transport from univalence and pole-skipping}},
  \href{http://dx.doi.org/10.1103/PhysRevLett.126.051601}{\emph{Phys. Rev.
  Lett.} {\bf 126} (2021) 051601}, [\href{http://arxiv.org/abs/2008.00888}{{\tt
  2008.00888}}].

\bibitem{Choi:2020tdj}
C.~Choi, M.~Mezei and G.~S\'arosi, \emph{{Pole skipping away from maximal
  chaos}},  \href{http://arxiv.org/abs/2010.08558}{{\tt 2010.08558}}.

\bibitem{Arean:2020eus}
D.~Arean, R.~A. Davison, B.~Gout\'eraux and K.~Suzuki, \emph{{Hydrodynamic
  diffusion and its breakdown near AdS$_2$ fixed points}},
  \href{http://arxiv.org/abs/2011.12301}{{\tt 2011.12301}}.

\bibitem{Wu:2021mkk}
N.~Wu, M.~Baggioli and W.-J. Li, \emph{{On the universality of AdS$_2$
  diffusion bounds and the breakdown of linearized hydrodynamics}},
  \href{http://arxiv.org/abs/2102.05810}{{\tt 2102.05810}}.

\bibitem{Grozdanov:2021gzh}
S.~Grozdanov, A.~Starinets and P.~Tadi\'c, \emph{{Hydrodynamic dispersion
  relations at finite coupling}},  \href{http://arxiv.org/abs/2104.11035}{{\tt
  2104.11035}}.

\bibitem{Maldacena:1997re}
J.~M. Maldacena, \emph{{The Large N limit of superconformal field theories and
  supergravity}}, \href{http://dx.doi.org/10.1023/A:1026654312961,
  10.1023/A:1026654312961}{\emph{Adv.Theor.Math.Phys.} {\bf 2} (1998)
  231--252}, [\href{http://arxiv.org/abs/hep-th/9711200}{{\tt
  hep-th/9711200}}].

\bibitem{Witten:1998qj}
E.~Witten, \emph{{Anti-de Sitter space and holography}}, {\emph{Adv. Theor.
  Math. Phys.} {\bf 2} (1998) 253--291},
  [\href{http://arxiv.org/abs/hep-th/9802150}{{\tt hep-th/9802150}}].

\bibitem{Gubser:1998bc}
S.~S. Gubser, I.~R. Klebanov and A.~M. Polyakov, \emph{{Gauge theory
  correlators from non-critical string theory}},
  \href{http://dx.doi.org/10.1016/S0370-2693(98)00377-3}{\emph{Phys. Lett.}
  {\bf B428} (1998) 105--114}, [\href{http://arxiv.org/abs/hep-th/9802109}{{\tt
  hep-th/9802109}}].

\bibitem{Buchbinder:2008dc}
E.~I. Buchbinder, S.~E. Vazquez and A.~Buchel, \emph{{Sound Waves in (2+1)
  Dimensional Holographic Magnetic Fluids}},
  \href{http://dx.doi.org/10.1088/1126-6708/2008/12/090}{\emph{JHEP} {\bf 12}
  (2008) 090}, [\href{http://arxiv.org/abs/0810.4094}{{\tt 0810.4094}}].

\bibitem{Buchbinder:2009aa}
E.~I. Buchbinder, \emph{Fate of sound and diffusion in a holographic magnetic
  field}, \href{http://dx.doi.org/10.1103/PhysRevD.79.046006}{\emph{Physical
  Review D} {\bf 79} (2009) }.

\bibitem{Hansen:2009xe}
J.~Hansen and P.~Kraus, \emph{{S-duality in AdS/CFT magnetohydrodynamics}},
  \href{http://dx.doi.org/10.1088/1126-6708/2009/10/047}{\emph{JHEP} {\bf 10}
  (2009) 047}, [\href{http://arxiv.org/abs/0907.2739}{{\tt 0907.2739}}].

\bibitem{Buchbinder:2009mk}
E.~I. Buchbinder and A.~Buchel, \emph{{Relativistic Conformal
  Magneto-Hydrodynamics from Holography}},
  \href{http://dx.doi.org/10.1016/j.physletb.2009.06.003}{\emph{Phys. Lett. B}
  {\bf 678} (2009) 135--138}, [\href{http://arxiv.org/abs/0902.3170}{{\tt
  0902.3170}}].

\bibitem{Hartnoll:2007ip}
S.~A. Hartnoll and C.~P. Herzog, \emph{{Ohm's Law at strong coupling: S duality
  and the cyclotron resonance}},
  \href{http://dx.doi.org/10.1103/PhysRevD.76.106012}{\emph{Phys.Rev.} {\bf
  D76} (2007) 106012}, [\href{http://arxiv.org/abs/0706.3228}{{\tt
  0706.3228}}].

\bibitem{Hansen:2008tq}
J.~Hansen and P.~Kraus, \emph{{Nonlinear Magnetohydrodynamics from Gravity}},
  \href{http://dx.doi.org/10.1088/1126-6708/2009/04/048}{\emph{JHEP} {\bf 04}
  (2009) 048}, [\href{http://arxiv.org/abs/0811.3468}{{\tt 0811.3468}}].

\bibitem{Hartnoll:2007ih}
S.~A. Hartnoll, P.~K. Kovtun, M.~Muller and S.~Sachdev, \emph{{Theory of the
  Nernst effect near quantum phase transitions in condensed matter, and in
  dyonic black holes}},
  \href{http://dx.doi.org/10.1103/PhysRevB.76.144502}{\emph{Phys.Rev.} {\bf
  B76} (2007) 144502}, [\href{http://arxiv.org/abs/0706.3215}{{\tt
  0706.3215}}].

\bibitem{Hernandez:2017mch}
J.~Hernandez and P.~Kovtun, \emph{{Relativistic magnetohydrodynamics}},
  \href{http://dx.doi.org/10.1007/JHEP05(2017)001}{\emph{JHEP} {\bf 05} (2017)
  001}, [\href{http://arxiv.org/abs/1703.08757}{{\tt 1703.08757}}].




\bibitem{Baggioli:2021ujk}
M.~Baggioli, U.~Gran and M.~Torns\"o, \emph{{Collective modes of polarizable
  holographic media in magnetic fields}},
  \href{http://arxiv.org/abs/2102.09969}{{\tt 2102.09969}}.
  
  
  
  
  
\bibitem{Amoretti:2021fch}
A.~Amoretti, D.~Arean, D.~K. Brattan and N.~Magnoli, \emph{{Hydrodynamic
  magneto-transport in charge density wave states}},
  \href{http://arxiv.org/abs/2101.05343}{{\tt 2101.05343}}.

\bibitem{Amoretti:2020mkp}
A.~Amoretti, D.~K. Brattan, N.~Magnoli and M.~Scanavino, \emph{{Magneto-thermal
  transport implies an incoherent Hall conductivity}},
  \href{http://dx.doi.org/10.1007/JHEP08(2020)097}{\emph{JHEP} {\bf 08} (2020)
  097}, [\href{http://arxiv.org/abs/2005.09662}{{\tt 2005.09662}}].

\bibitem{Amoretti:2019buu}
A.~Amoretti, M.~Meinero, D.~K. Brattan, F.~Caglieris, E.~Giannini, M.~Affronte
  et~al., \emph{{Hydrodynamical description for magneto-transport in the
  strange metal phase of Bi-2201}},
  \href{http://dx.doi.org/10.1103/PhysRevResearch.2.023387}{\emph{Phys. Rev.
  Res.} {\bf 2} (2020) 023387}, [\href{http://arxiv.org/abs/1909.07991}{{\tt
  1909.07991}}].

\bibitem{Kim:2015wba}
K.-Y. Kim, K.~K. Kim, Y.~Seo and S.-J. Sin, \emph{{Thermoelectric
  Conductivities at Finite Magnetic Field and the Nernst Effect}},
  \href{http://dx.doi.org/10.1007/JHEP07(2015)027}{\emph{JHEP} {\bf 07} (2015)
  027}, [\href{http://arxiv.org/abs/1502.05386}{{\tt 1502.05386}}].

\bibitem{Jeong:2021zhz}
H.-S. Jeong, K.-Y. Kim and Y.-W. Sun, \emph{{Bound of diffusion constants from
  pole-skipping points: spontaneous symmetry breaking and magnetic field}},
  \href{http://arxiv.org/abs/2104.13084}{{\tt 2104.13084}}.

\bibitem{Blake:2015hxa}
M.~Blake, \emph{{Magnetotransport from the fluid/gravity correspondence}},
  \href{http://dx.doi.org/10.1007/JHEP10(2015)078}{\emph{JHEP} {\bf 10} (2015)
  078}, [\href{http://arxiv.org/abs/1507.04870}{{\tt 1507.04870}}].

\bibitem{Li:2019bgc}
W.~Li, S.~Lin and J.~Mei, \emph{{Thermal diffusion and quantum chaos in neutral
  magnetized plasma}},
  \href{http://dx.doi.org/10.1103/PhysRevD.100.046012}{\emph{Phys. Rev. D} {\bf
  100} (2019) 046012}, [\href{http://arxiv.org/abs/1905.07684}{{\tt
  1905.07684}}].

\bibitem{Kaminski:2009dh}
M.~Kaminski, K.~Landsteiner, J.~Mas, J.~P. Shock and J.~Tarrio,
  \emph{{Holographic Operator Mixing and Quasinormal Modes on the Brane}},
  \href{http://dx.doi.org/10.1007/JHEP02(2010)021}{\emph{JHEP} {\bf 1002}
  (2010) 021}, [\href{http://arxiv.org/abs/0911.3610}{{\tt 0911.3610}}].

\bibitem{Faulkner:2009wj}
T.~Faulkner, H.~Liu, J.~McGreevy and D.~Vegh, \emph{{Emergent quantum
  criticality, Fermi surfaces, and AdS(2)}},
  \href{http://dx.doi.org/10.1103/PhysRevD.83.125002}{\emph{Phys.Rev.} {\bf
  D83} (2011) 125002}, [\href{http://arxiv.org/abs/0907.2694}{{\tt
  0907.2694}}].

\bibitem{Hartnoll:2012rj}
S.~A. Hartnoll and D.~M. Hofman, \emph{{Locally Critical Resistivities from
  Umklapp Scattering}},
  \href{http://dx.doi.org/10.1103/PhysRevLett.108.241601}{\emph{Phys.Rev.Lett.}
  {\bf 108} (2012) 241601}, [\href{http://arxiv.org/abs/1201.3917}{{\tt
  1201.3917}}].

\bibitem{Blake:2016sud}
M.~Blake, \emph{{Universal Diffusion in Incoherent Black Holes}},
  \href{http://dx.doi.org/10.1103/PhysRevD.94.086014}{\emph{Phys. Rev.} {\bf
  D94} (2016) 086014}, [\href{http://arxiv.org/abs/1604.01754}{{\tt
  1604.01754}}].

\bibitem{Blake:2017qgd}
M.~Blake, R.~A. Davison and S.~Sachdev, \emph{{Thermal diffusivity and chaos in
  metals without quasiparticles}},  \href{http://arxiv.org/abs/1705.07896}{{\tt
  1705.07896}}.

\bibitem{Blake:2016jnn}
M.~Blake and A.~Donos, \emph{{Diffusion and Chaos from near AdS$_2$ horizons}},
  \href{http://dx.doi.org/10.1007/JHEP02(2017)013}{\emph{JHEP} {\bf 02} (2017)
  013}, [\href{http://arxiv.org/abs/1611.09380}{{\tt 1611.09380}}].

\bibitem{Baggioli:2017ojd}
M.~Baggioli and W.-J. Li, \emph{{Diffusivities bounds and chaos in holographic
  Horndeski theories}},
  \href{http://dx.doi.org/10.1007/JHEP07(2017)055}{\emph{JHEP} {\bf 07} (2017)
  055}, [\href{http://arxiv.org/abs/1705.01766}{{\tt 1705.01766}}].

\bibitem{Kim:2017dgz}
K.-Y. Kim and C.~Niu, \emph{{Diffusion and Butterfly Velocity at Finite
  Density}}, \href{http://dx.doi.org/10.1007/JHEP06(2017)030}{\emph{JHEP} {\bf
  06} (2017) 030}, [\href{http://arxiv.org/abs/1704.00947}{{\tt 1704.00947}}].

\bibitem{Ahn:2017kvc}
H.-S. Jeong, Y.~Ahn, D.~Ahn, C.~Niu, W.-J. Li and K.-Y. Kim, \emph{{Thermal
  diffusivity and butterfly velocity in anisotropic Q-Lattice models}},
  \href{http://dx.doi.org/10.1007/JHEP01(2018)140}{\emph{JHEP} {\bf 01} (2018)
  140}, [\href{http://arxiv.org/abs/1708.08822}{{\tt 1708.08822}}].

\bibitem{Grozdanov:2017ajz}
S.~Grozdanov, K.~Schalm and V.~Scopelliti, \emph{{Black hole scrambling from
  hydrodynamics}},
  \href{http://dx.doi.org/10.1103/PhysRevLett.120.231601}{\emph{Phys. Rev.
  Lett.} {\bf 120} (2018) 231601}, [\href{http://arxiv.org/abs/1710.00921}{{\tt
  1710.00921}}].

\bibitem{larkin1969quasiclassical}
A.~Larkin and Y.~N. Ovchinnikov, \emph{{Quasiclassical method in the theory of
  superconductivity}}, {\emph{Sov Phys JETP} (1969) }.

\bibitem{Kitaev-2014}
A.~Kitaev, \emph{{A simple model of quantum holography}}, {\emph{Talks at KITP}
  (April 7, 2015 and May 27, (2015))
  http://online.kitp.ucsb.edu/online/entangled15/kitaev/,
  http://online.kitp.ucsb.edu/online/entangled15/kitaev2/}.

\bibitem{Maldacena:2015waa}
J.~Maldacena, S.~H. Shenker and D.~Stanford, \emph{{A bound on chaos}},
  \href{http://dx.doi.org/10.1007/JHEP08(2016)106}{\emph{JHEP} {\bf 08} (2016)
  106}, [\href{http://arxiv.org/abs/1503.01409}{{\tt 1503.01409}}].

\bibitem{Baggioli:2021xuv}
M.~Baggioli, K.-Y. Kim, L.~Li and W.-J. Li, \emph{{Holographic Axion Model: a
  simple gravitational tool for quantum matter}},
  \href{http://arxiv.org/abs/2101.01892}{{\tt 2101.01892}}.

\bibitem{Baggioli:2020ljz}
M.~Baggioli and W.-J. Li, \emph{{Universal Bounds on Transport in Holographic
  Systems with Broken Translations}},
  \href{http://dx.doi.org/10.21468/SciPostPhys.9.1.007}{\emph{SciPost Phys.}
  {\bf 9} (2020) 007}, [\href{http://arxiv.org/abs/2005.06482}{{\tt
  2005.06482}}].

\bibitem{Blake:2018leo}
M.~Blake, R.~A. Davison, S.~Grozdanov and H.~Liu, \emph{{Many-body chaos and
  energy dynamics in holography}},
  \href{http://dx.doi.org/10.1007/JHEP10(2018)035}{\emph{JHEP} {\bf 10} (2018)
  035}, [\href{http://arxiv.org/abs/1809.01169}{{\tt 1809.01169}}].
  

\end{thebibliography}
\bibliographystyle{JHEP}

\providecommand{\href}[2]{#2}\begingroup\raggedright\endgroup

\end{document}